\documentclass[manuscript,screen,sigconf,10pt,nonacm]{acmart}
\AtBeginDocument{%
  }

\usepackage{xcolor}
\usepackage{subcaption}
\usepackage{xspace}
\usepackage{mhchem}
\usepackage{pifont}
\newcommand{\cname}{\emph{Solar-CSK}\xspace}
\definecolor{Change}{RGB}{0,0,255}

\setcopyright{acmlicensed}
\copyrightyear{2018}
\acmYear{2018}
\acmDOI{XXXXXXX.XXXXXXX}
\acmConference[Conference acronym 'XX]{Make sure to enter the correct
  conference title from your rights confirmation email}{June 03--05,
  2018}{Woodstock, NY}

\setcopyright{none}
\acmISBN{978-1-4503-XXXX-X/2018/06}

\begin{document}

\title{Solar-CSK: Decoding Color Coded Visible Light Communications using Solar Cells}

\author{Yanxiang Wang$^{1,2}$,\hspace{3pt} Yihe Yan$^{1}$,\hspace{3pt} Jiawei Hu$^{1,2}$,\hspace{3pt} Cheng Jiang$^1$, \\\hspace{3pt} Brano Kusy$^2$, \hspace{3pt} Ashraf Uddin$^1$, \hspace{3pt} Mahbub Hassan$^1$,\hspace{3pt} Wen Hu$^1$}

\affiliation{
    \institution{$^1$ University of New South Wales, Australia\\
    $^2$ Data61-CSIRO, Australia \\
    \country{\tt\small Email:~\{yanxiang.wang, yihe.yan, jiawei.hu\}@unsw.edu.au,  cheng.jiang1@student.unsw.edu.au \\ \{a.uddin, mahbub.hassan, wen.hu\}@unsw.edu.au,  brano.kusy@data61.csiro.au}
}}
\renewcommand{\authors}{Yanxiang Wang, Jiawei Hu, Hong Jia, Wen Hu, Mahbub Hassan, Ashraf Uddin, Brano Kusy, Moustafa Youssef}









\renewcommand{\shortauthors}{Wang et al.}

\begin{abstract}
Visible Light Communication (VLC) provides an energy-efficient wireless solution by using existing LED-based illumination for high-speed data transmissions. Although solar cells offer the advantage of simultaneous energy harvesting and data reception, their broadband nature hinders accurate decoding of color-coded signals like Color Shift Keying (CSK). In this paper, we propose a novel approach exploiting the concept of tandem solar cells, multi-layer devices with partial wavelength selectivity, to capture coarse color information without resorting to energy-limiting color filters. To address the residual spectral overlap, we develop a bidirectional LSTM-based machine learning framework that infers channel characteristics by comparing solar cells' photovoltaic signals with pilot-based anchor data. Our commercial off-the-shelf (COTS) solar prototype achieves robust performance across varying distances and ambient lighting levels, significantly reducing bit error rates compared to conventional channel estimation methods. These findings mark a step toward sustainable, high-performance VLC systems powered by the multi-layer solar technologies.
\end{abstract}

\maketitle

\section{Introduction}

Visible Light Communication (VLC) has emerged as a promising wireless solution by leveraging energy-efficient LED lighting for high-speed data transmissions~\cite{grobe2013high,james2020micro,wang2014high,li2014550}. Traditionally, photodiodes have dominated VLC receivers~\cite{xu2023visible,xu2022low,ghiasi2024exploiting,xu2022exploiting,ye2023vlc,li2016practical}; however, solar cells have recently gained attention for their ability to both harvest energy and decode data simultaneously~\cite{sarwar2017visible,fakidis2020simultaneous,de2022self,kadirvelu2021circuit,wu20180,das2019towards}. While solar-based VLC excels in simple modulation schemes such as On-Off Keying (OOK), it faces a significant challenge with color-coded signaling, e.g., Color Shift Keying (CSK). Standard silicon solar cells are inherently broadband and cannot distinguish between different wavelengths, and using color filters to separate red, green, and blue channels drastically reduces energy-harvesting efficiency—negating the main advantage of sustainability.

To address this limitation, we propose exploiting \textit{tandem}~\cite{tandem} solar cells, a new class of multi-layer (or multi-junction) devices designed to capture a broader portion of the visible (and sometimes infrared) spectrum through layers with different bandgaps~\cite{tandem,bett1999iii,yamaguchi2018review,snaith2013perovskites}. By splitting incoming light across these specialized layers made from different materials, tandem cells minimize energy losses that occur when photons do not match a single bandgap, thus significantly boosting overall power conversion efficiency~\cite{umeno1991first,essig2015progress,essig2016realization}. This layered design also offers partial color discrimination—each layer is more responsive to certain wavelengths—making tandem solar cells better suited than traditional single-junction devices for color-coded VLC applications.
\begin{figure}[t!]
    \centering
    \includegraphics[width=0.9\linewidth]{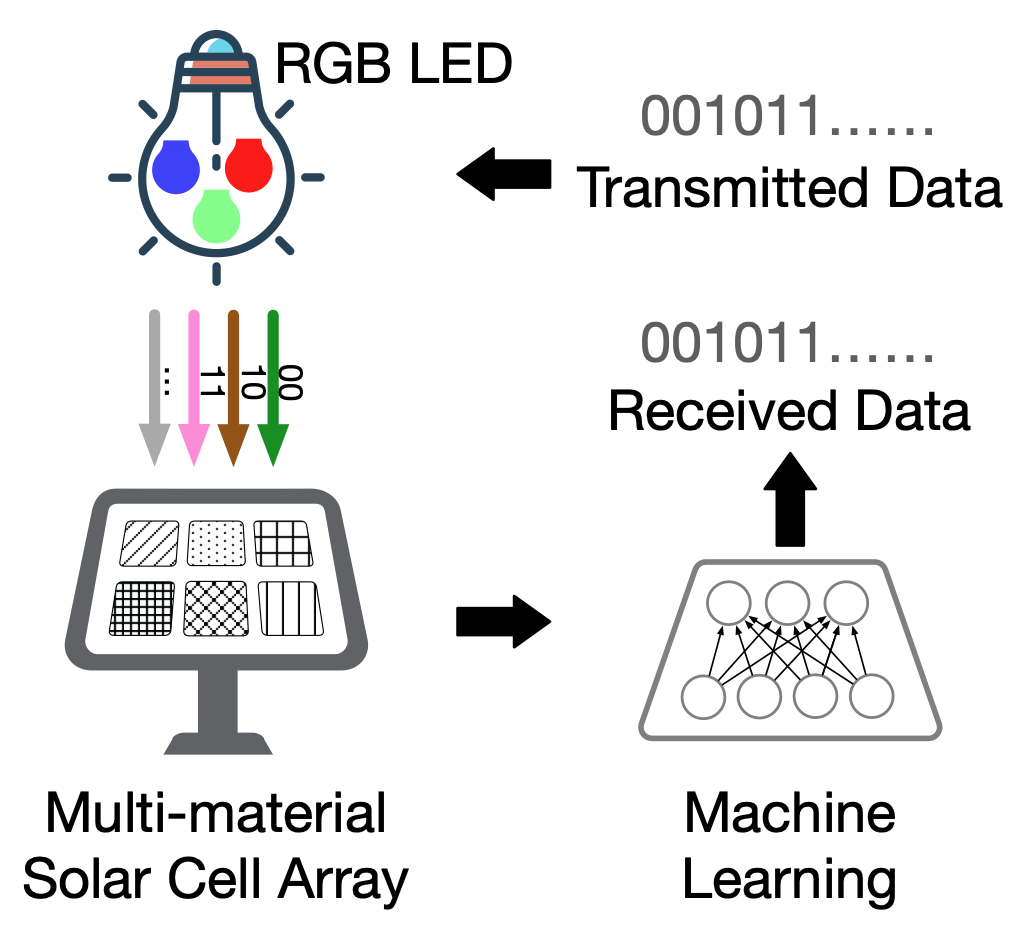}
    \caption{A multi-material solar array decodes CSK signals transmitted by RGB LEDs using machine learning.}
    \label{fig:teaser}
\end{figure}
However, the layers or sub-cells of a tandem solar cell are not narrowband filters, meaning their spectral overlap still yields only \textit{coarse} color information that complicates traditional channel estimation for CSK that use photodiodes with integrated color filters. In this work, we develop a machine learning approach—specifically, a bidirectional long short-term memory (LSTM) model with differential solar energy inputs—where the solar Analog-to-Digital Converter (ADC) readings are subtracted from pilot-based anchor data to infer channel characteristics. Our prototype (illustrated in Fig.~\ref{fig:teaser}), built with commercial off-the-shelf (COTS) solar cells composed of different materials, demonstrates superior performance in decoding color-coded VLC compared to conventional methods, moving us closer to sustainable, reliable light-based wireless communications.

Our key contributions can be summarized as follows:
\begin{itemize}
    \item To the best of our knowledge, we present the first CSK decoding approach leveraging the concept of tandem solar cells, without resorting to color filters, by exploiting their multi-layer design to achieve partial color discrimination.
    \item We demonstrate that the partial color discrimination provided by tandem solar cells is insufficient for conventional channel estimation and calibration methods to handle unseen channels. (\S \ref{s:eval}) 
    \item We present a novel end-to-end machine‑learning approach for CSK decoding with tandem solar cells, directly recovering transmitted colors from solar cell outputs. (\S \ref{s:syst-model})
    \item Compared to conventional channel estimation and decoding, our COTS solar prototype significantly reduces bit error rates, delivering robust performance across varying distances and ambient lighting. (\S \ref{s:eval})
\end{itemize}

\section{Background}
\subsection{CIE 1931 Color Space}
\begin{figure}[htp!]
\begin{subfigure}[b]{0.44\columnwidth}
\centering
\includegraphics[width=\textwidth]{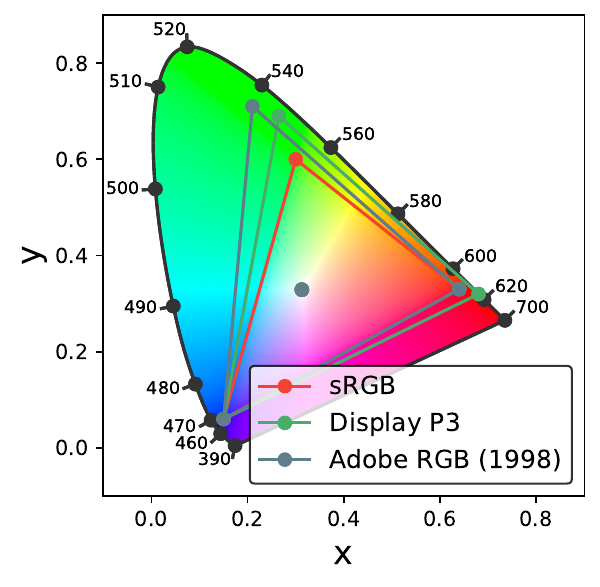}
\caption{}
\end{subfigure}
\hfill
\begin{subfigure}[b]{0.545\columnwidth}
\centering
\includegraphics[width=\textwidth]{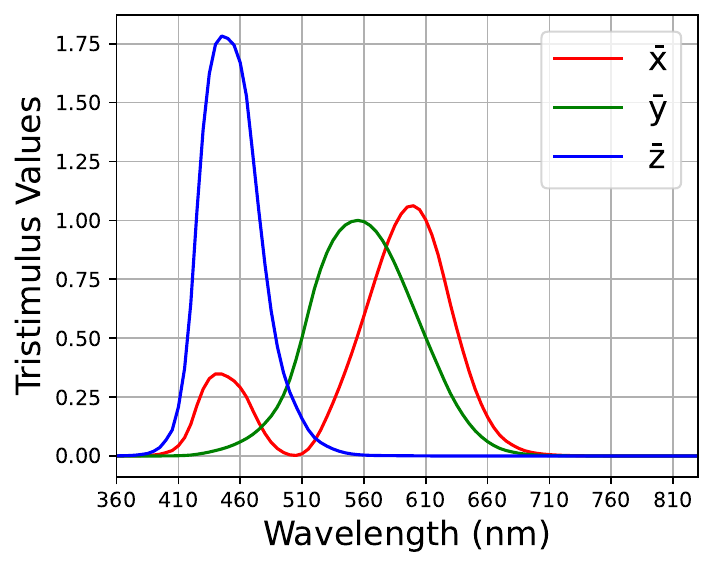}
\caption{}
\end{subfigure}
\caption{(a) CIE color space. (b) CIE XYZ standard observer color matching functions.}
\label{fig:color-space}
\end{figure}

The relationship between the wavelength of light and color perceived by humans is defined in the CIE\footnote{CIE stands for ``Commission Internationale de l'éclairage'' (International Commission on Illumination)} color space~\cite{smith1931cie}, as shown in Fig.~\ref{fig:color-space} (a). This color space has two dimensions, $x$ and $y$, where $x=\frac{X}{X+Y+Z}$ and $y=\frac{Y}{X+Y+Z}$. The tristimulus values $X$, $Y$, and $Z$ for a color can be calculated from spectral distributions by applying the standard color matching functions~\cite{wyszecki2000color}, as illustrated in Fig.~\ref{fig:color-space} (b). Other color spaces, such as Adobe RGB (1998), sRGB, and Display P3, are used for various applications, including print production, professional photography, and display on screens. It should be noted that a color can be mapped to either a single wavelength, for example, red (around 700 nm), or a combination of light at different wavelengths, for instance, white is a combination of all visible wavelengths. Black, on the other hand, is the absence of light.
\subsection{Color-Shift Keying Modulation}
\begin{figure}[htp!]
\begin{subfigure}[b]{0.32\columnwidth}
\centering
\includegraphics[width=\textwidth]{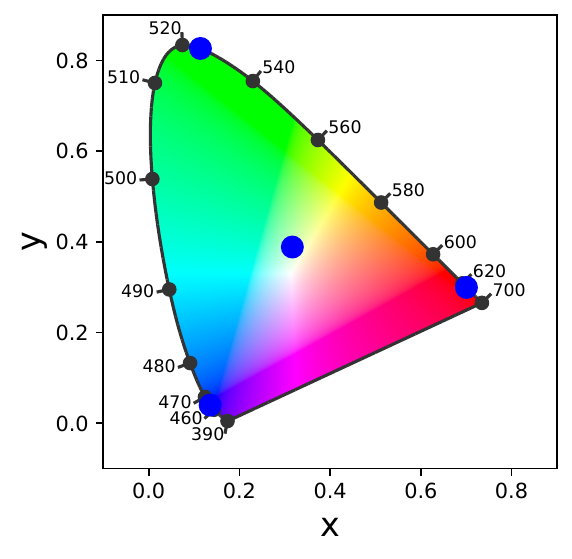}
\caption{4-CSK}
\end{subfigure}
\hfill
\begin{subfigure}[b]{0.32\columnwidth}
\centering
\includegraphics[width=\textwidth]{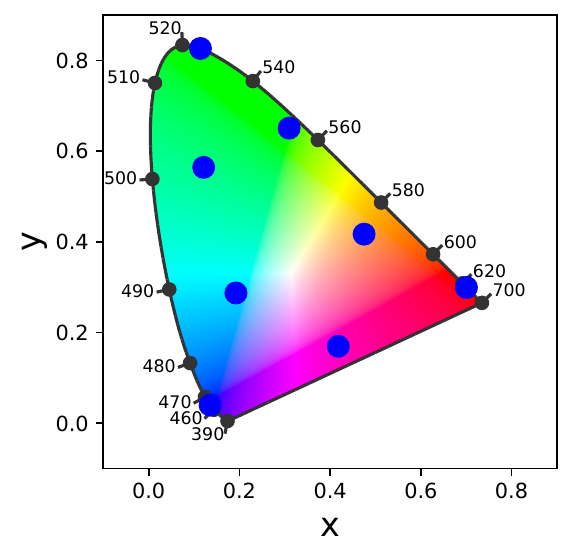}
\caption{8-CSK}
\end{subfigure}
\hfill
\begin{subfigure}[b]{0.32\columnwidth}
\centering
\includegraphics[width=\textwidth]{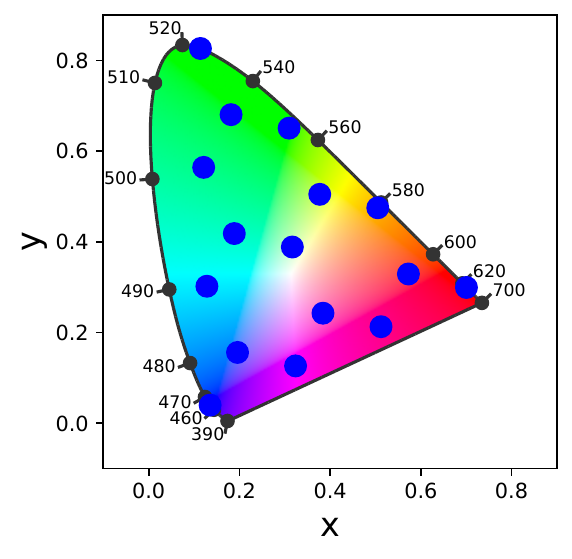}
\caption{16-CSK}
\end{subfigure}
\caption{IEEE 802.15.7 CSK constellations (blue dots).}
\label{fig:constellation}
\end{figure}

Color-Shift Keying (CSK) modulation, introduced in the IEEE 802.15.7 standard~\cite{ieee2011ieee}, was designed for short-range optical wireless communications. Unlike intensity-based modulation schemes, CSK maintains a constant power envelope across transmitted symbols, thereby mitigating potential health risks associated with light intensity fluctuations~\cite{monteiro2014design}.
The standard allows us to transmit data by changing the color of light. We can select three distinct colors from seven available color bands within the visible spectrum~\cite{ieee2011ieee}. For simplicity, we use Red, Green, and Blue and plot these colors on a standard CIE chart, where they form a triangle (see Fig.~\ref{fig:constellation}). We use this triangular color space to encode data. In the simplest case, we place three of the color anchors at the corners of the triangle and one in the center, representing four symbols. The standard also permits more complex setups with eight or sixteen color anchors, which allows us to send more data at once, as shown in Fig.~\ref{fig:constellation} (b) and (c). However, higher number of symbols introduce a trade-off as individual symbols interfere with each other more, potentially leading to more errors.

\begin{figure*}[t]
    \centering
    \begin{minipage}[t]{0.32\linewidth}
        \centering
        \includegraphics[width=0.85\textwidth]{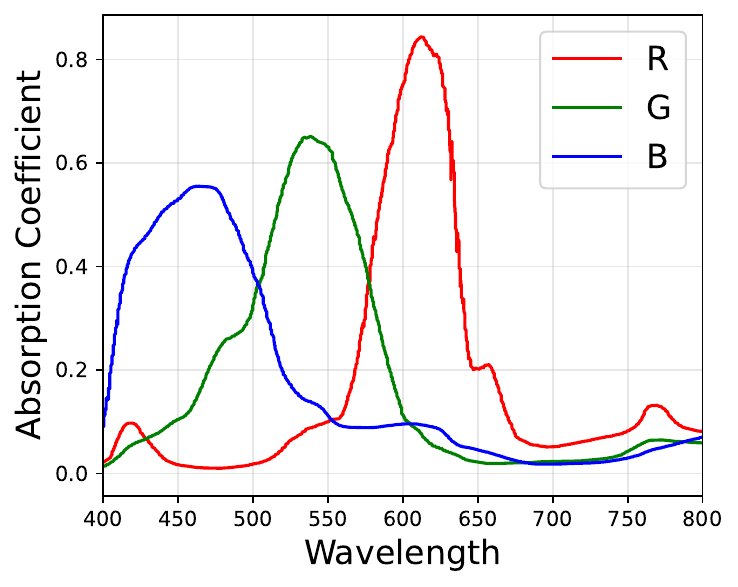}
        \caption{Spectral response of the true color sensor AS73211.}
        \label{fig:true-color}
    \end{minipage}%
    \hfill%
    \begin{minipage}[t]{0.32\linewidth}
        \centering
        \includegraphics[width=0.85\textwidth]{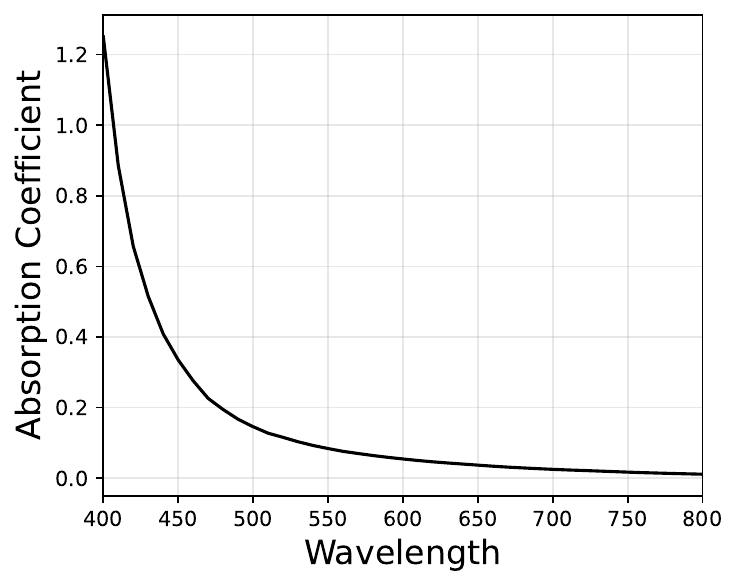}
        \caption{Spectral response of a silicon-based solar cell.}
        \label{fig:silicon}
    \end{minipage}%
    \hfill%
    \begin{minipage}[t]{0.32\linewidth}
        \centering
        \includegraphics[width=0.85\textwidth]{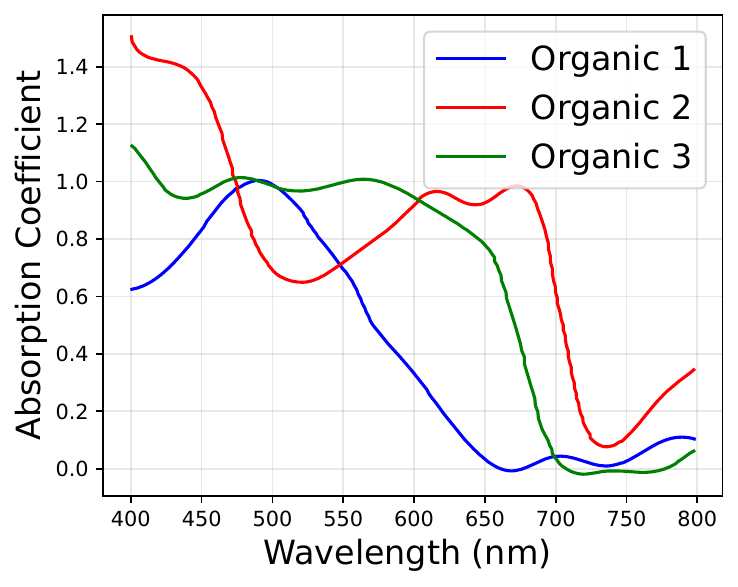}
        \caption{Spectral response of three distinct-material organic solar cells}
        \label{fig:diff-absorption}
    \end{minipage}
\end{figure*}
\subsection{True Color Sensor vs Silicon Solar cell}
For CSK modulation, true color sensors are commonly used as receivers due to their accurate spectral response that is very close to the CIE color matching function (see  Fig.~\ref{fig:color-space}b). This response is achieved by adjusting the color filters positioned on the top of photo-diodes. Solar cells, while also photovoltaic devices, primarily function for energy harvesting and operate in non-biased mode, making them less sensitive than photo-diodes. More significantly, solar cells have a wider absorption range and cannot resolve signals into narrow spectral bands. As a result, they output a single accumulated or scalar value for different color bands. For instance, Fig.~\ref{fig:true-color}  and Fig.~\ref{fig:silicon} present the spectral responses of the true color sensor AS73211~\cite{73211} and silicon-based solar cells~\cite{green1995optical}. While the true color sensor has three channels, each with peaks at different wavelengths, the silicon based solar cell exhibit a monotonic behavior with decreasing absorption capabilities as the wavelength increases.

\subsection{Tandem Solar Cells}
To enable solar cells to function as receivers for CSK modulation, we propose utilizing tandem solar cells~\cite{tandem} instead of applying external physical color filters, which could compromise energy harvesting efficiency and add extra cost. 

Tandem solar cells use  multi-junction design that incorporates layers made from different materials (similar to the diverse array of solar cell materials previously considered) and naturally provides color discrimination capabilities. Each specialized layer within the tandem structure responds to a different wavelength in the visible spectrum, which makes the layers inherently suitable for color-based communication. Beyond traditional silicon, these layers can integrate various materials, including amorphous silicon and organic-based compounds. Notably, organic components in tandem cells retain flexibility and semi-transparency, enabling their application as gesture recognition sensors when affixed to device surfaces while simultaneously harvesting energy~\cite{ma2019solargest}. For simplicity, we implement our prototype using an array of solar cells fabricated from different materials, which offers similar spectral sensitivity benefits while allowing for more straightforward experimental validation.
\begin{figure}[htp!]
\begin{subfigure}[b]{0.49\columnwidth}
    \centering
    \includegraphics[width=\linewidth]{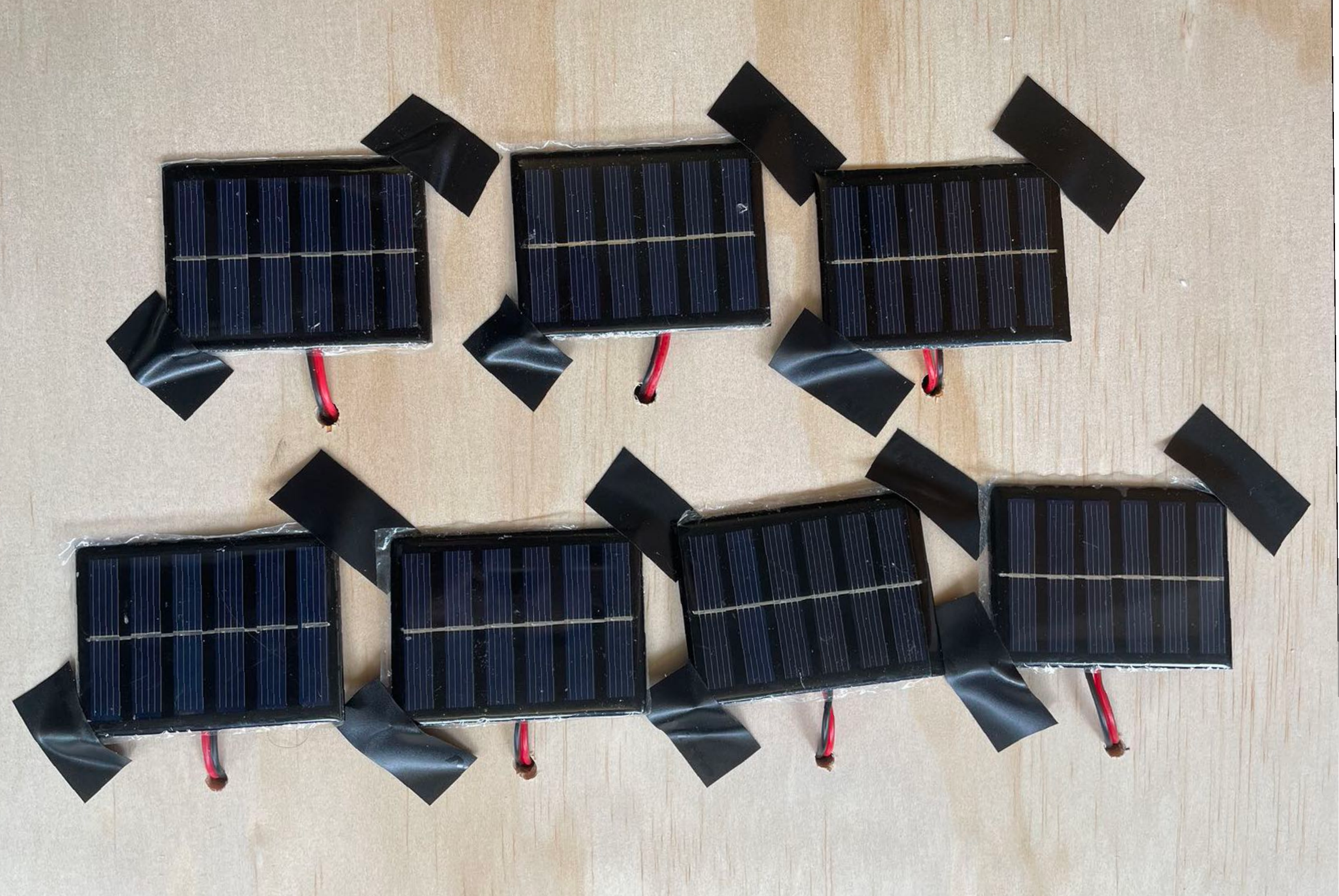}
    \caption{}
\end{subfigure}
    \hfill
\begin{subfigure}[b]{0.44\columnwidth}
    \centering
    \includegraphics[width=\linewidth]{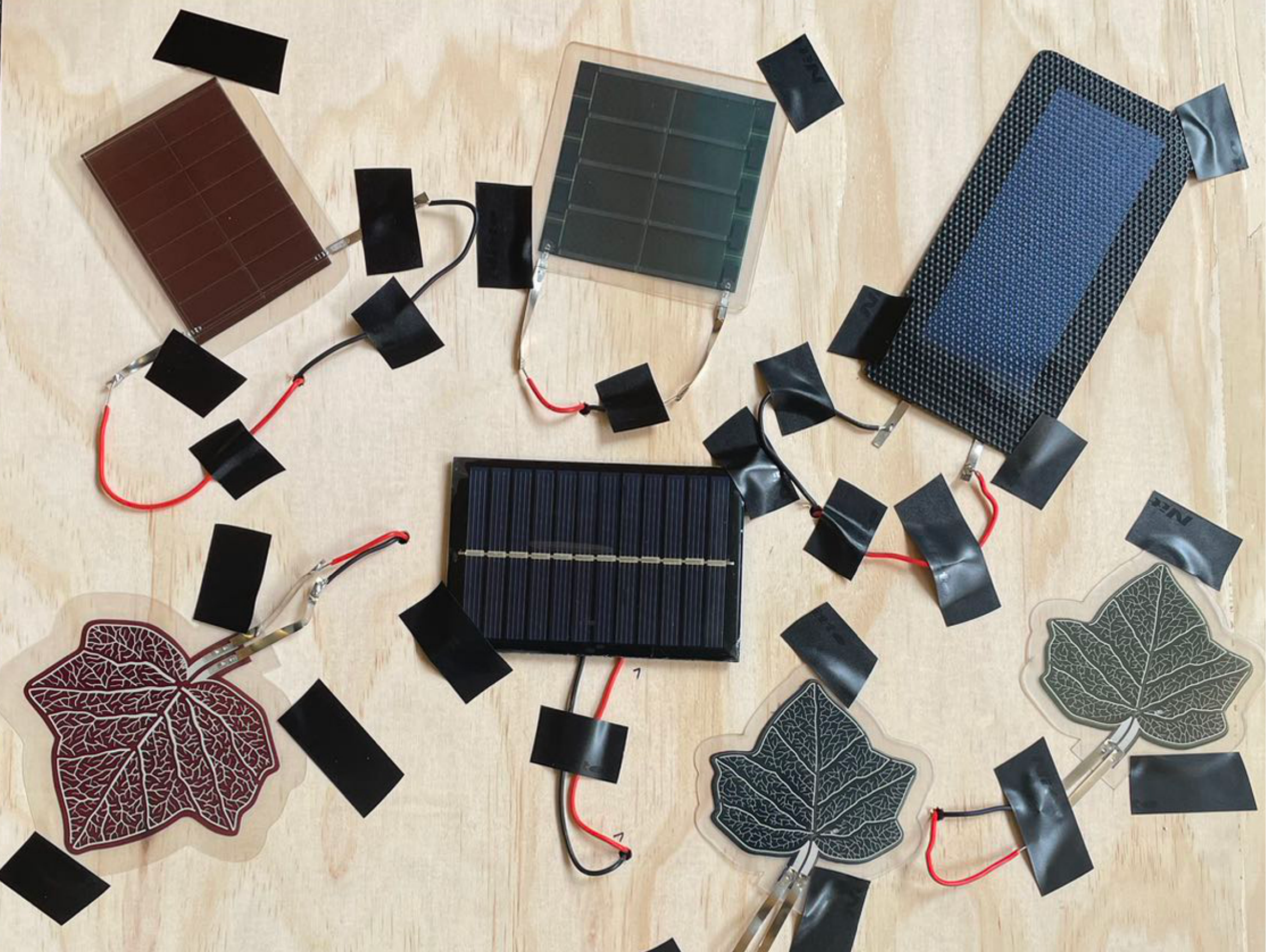}
    \caption{}
\end{subfigure}
    \caption{(a) Seven silicon solar cells; (b) Seven multi-material cells: one polycrystalline, one amorphous, and five organic.}
    \label{fig:array}
\end{figure}

Figure~\ref{fig:diff-absorption} displays the absorption coefficients of three organic solar cells at various wavelengths\footnote{The organic solar cells and their associated absorption coefficient data were acquired from ASCA Technology https://www.asca.com/asca-technology/.}. Each solar cell exhibits a characteristic absorption profile, with peaks and troughs at different wavelengths, reflecting its specific material properties and structure. This variability in absorption patterns across the spectrum is particularly noteworthy, as it suggests potential advantages in combining different cell types for broader spectral coverage. While absorption data offer valuable insights into cell optical properties, obtaining accurate measurements presents challenges. The packaging material can significantly alter the solar cells' optical characteristics. Furthermore, overlapping absorption curves complicate data analysis and signal decoding compared to true color sensor measurements.

\section{Color Separation with Tandem Cells}

\begin{figure}[htp!]
\begin{subfigure}[b]{0.49\columnwidth}
    \centering
    \includegraphics[width=\linewidth]{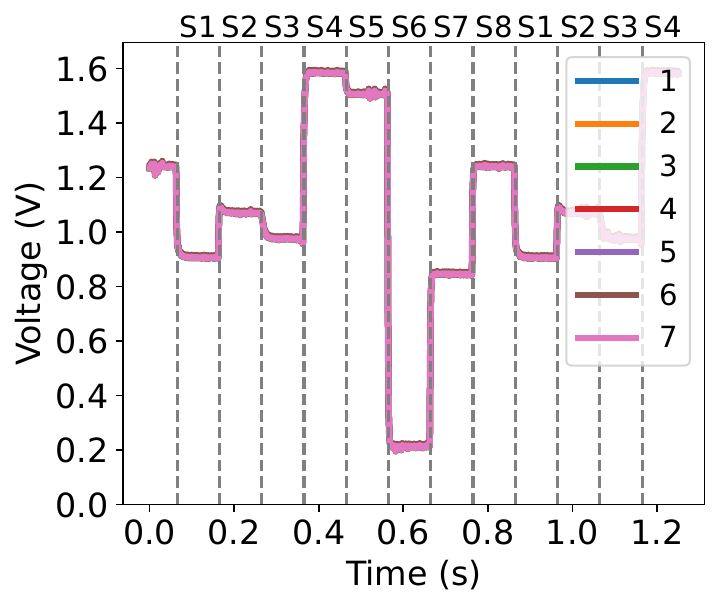}
    \caption{}
\end{subfigure}
    \hfill
\begin{subfigure}[b]{0.49\columnwidth}
    \centering
    \includegraphics[width=\linewidth]{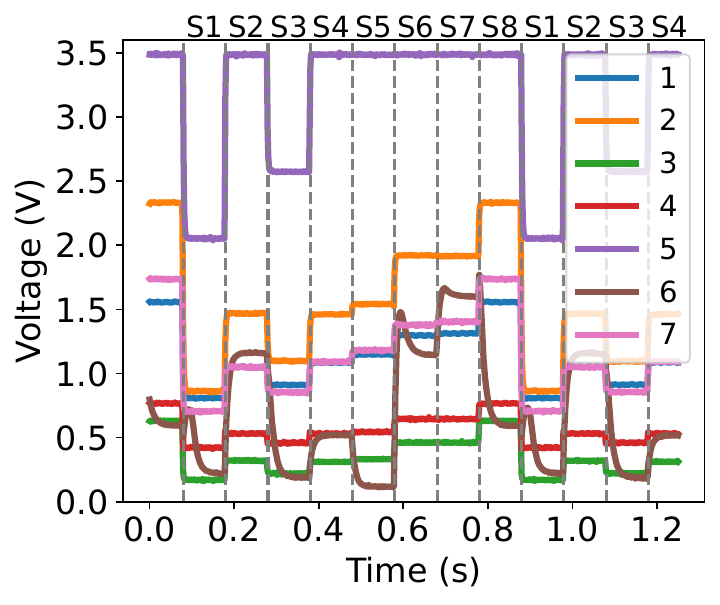}
    \caption{}
\end{subfigure}
    \caption{The time-domain voltage measurements from seven solar cells, (a) all silicon solar cells (b) each fabricated with different materials, over a 1.2-second interval. The individual curves represent the voltage output from each solar cell as the transmitted light color changes at a frequency of 10 Hz. Vertical dashed lines indicate the switching intervals between different color states.}
    \label{fig:time-domain}
\end{figure}

We further investigated the potential benefits of employing a solar cell array composed of multiple materials through a preliminary experiment. We used two solar cell arrays: one consisted of seven silicon solar cells, while the other was comprised of seven solar cells with different materials: one poly-crystalline silicon, one amorphous silicon, and five organic material-based solar cells, as shown in Fig.~\ref{fig:array}. To study the arrays' responses to pure, monochromatic light sources, we illuminated both the silicon cell array and the multi-material solar cell array with eight distinct colors (S1,S2,...,S8), which were acquired according to IEEE standards, and maintained consistent light power by measuring with a light meter. We alternated between them at a frequency of 10 Hz. Our observations in Fig.~\ref{fig:time-domain} revealed that different solar cells respond distinctly to identical stimuli, while the pure silicon solar cells behave similarly. For instance, in Fig.~\ref{fig:time-domain} (b), solar cells 1 and 2 show very similar behavior during the 0.1-0.2s interval but diverge significantly after 0.3s. Moreover, the response characteristics vary among cells, with solar cell 6 notably requiring a longer stabilization period following light changes. On the other hand, the data from the silicon cell array is essentially very similar; the readings overlap, and we cannot observe variance across different cells. 

\begin{figure}[htp!]
\begin{subfigure}[b]{0.49\columnwidth}
    \centering
    \includegraphics[width=\linewidth]{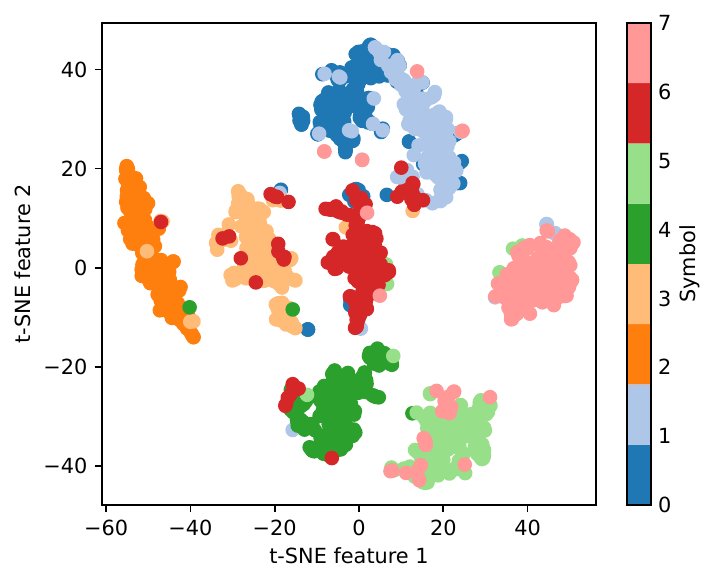}
    \caption{}
    \label{fig:rgb-response}
\end{subfigure}
    \hfill
\begin{subfigure}[b]{0.49\columnwidth}
    \centering
    \includegraphics[width=\linewidth]{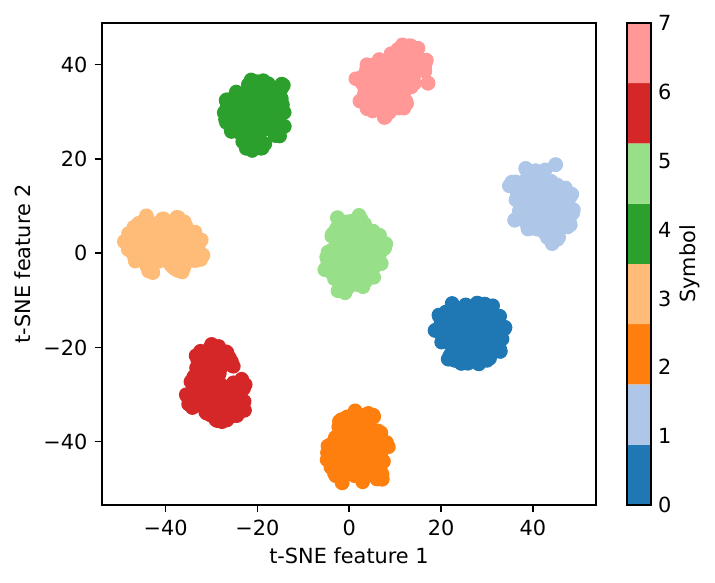}
    \caption{}
    \label{fig:rgb-response}
\end{subfigure}
    \caption{t-SNE visualization (a) seven silicon solar cells, (b) seven solar cells with different materials.}
    \label{fig:t-sne}
\end{figure}

Further more, using the photo-voltaic voltage measurements obtained from the solar cells as input, we visualize the t-SNE distributions in Fig.~\ref{fig:t-sne}. The results demonstrate that with a larger variety of solar cells, eight distinct symbol clusters emerge clearly in Fig.~\ref{fig:t-sne} (b), in contrast to the significant overlap and confusion observed in Fig.~\ref{fig:t-sne} (a) when using seven silicon solar cells. This result confirms the feasibility of separating colours using tandem solar cells.

However, implementing solar cell arrays as receivers for CSK presents a key challenge: extracting robust color features. Unlike true color sensors, solar cells operating in a non-biased mode cannot achieve comparable color response curves. This makes them more susceptible to channel variations, where factors such as distance and ambient light significantly affect voltage readings.  
To mitigate channel variability,
we propose utilizing \textit{differential measurements between incoming samples and predefined anchor symbols}. These anchor symbols are embedded within the preamble sequence to periodically capture channel background information. Since both data packet symbols and anchor symbols experience identical channel conditions, their relative differences remain consistent and robust against environmental fluctuations.

\section{\cname System Model}
\label{s:syst-model}
\begin{figure}[htp!]
    \centering
    \includegraphics[width=\linewidth]{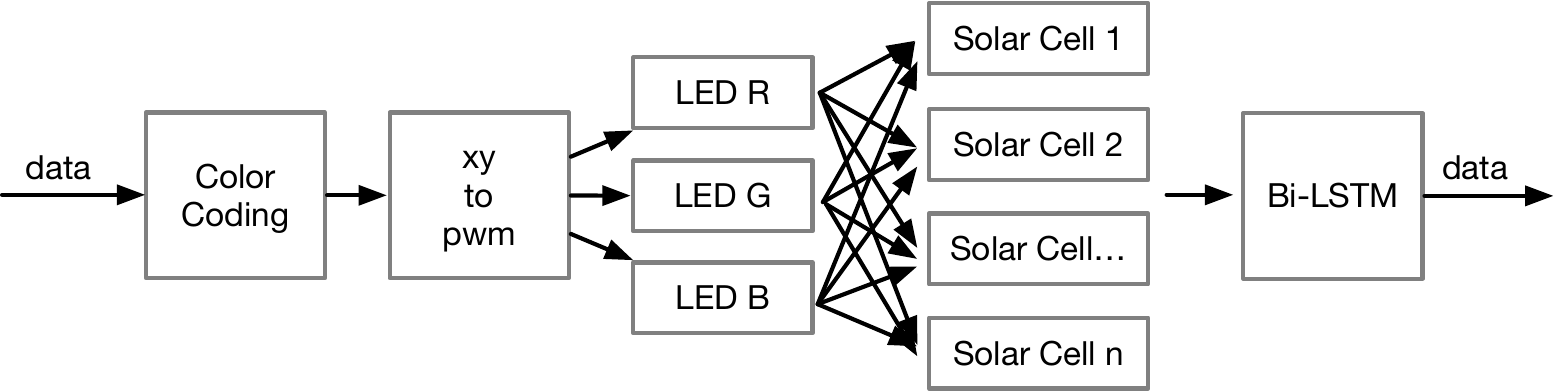}
    \caption{The pipeline of the proposed \cname.}
    \label{fig:pipeline}
\end{figure}

We consider a VLC system in which there is a transmitter and a receiver. The transmitter is an RGB LED with the diver circuit. The light emitted from the LED is modulated to be in different color or spectrum, representing different symbols. At the receiver, a solar cell array containing solar cell with different materials is utilized to capture the spectral information of the transmitted light signals. As illustrated in Fig.~\ref{fig:pipeline}, the data is encoded using different colors and converted to Pulse Width Modulation (PWM) values that control RGB LED power output. Photo-voltaic signals from solar cells undergo processing to extract features, which are then input to a demodulation network based on bidirectional LSTM architecture. The symbols are classified and ultimately converted back to their original data format.
\subsection{Transmitter}
\begin{figure}[htp!]
    \centering
    \includegraphics[width=0.6\linewidth]{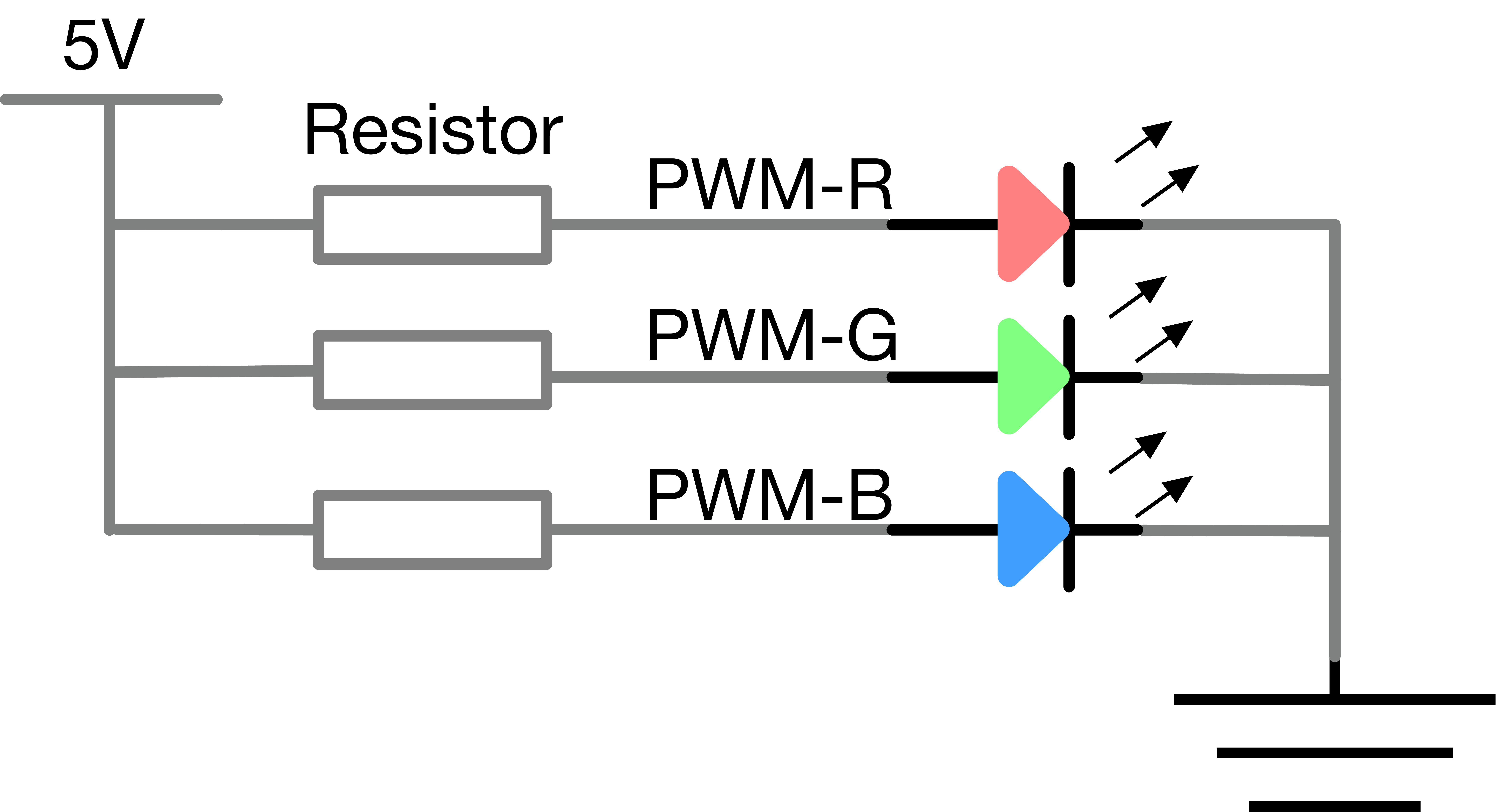}
    \caption{Circuit diagram of the transmitter.}
    \label{fig:transimitter}
\end{figure}
To achieve precise color control in lighting applications, RGB LEDs serve as the primary illumination source, featuring three distinct wavelength channels: red centered at 625 nm, green at 525 nm, and blue at 465 nm. In addition, RGB LEDs are also widely applied in smart lighting infrastructure, allowing for straightforward retrofitting and compatibility with modern lighting systems.

For precise color control, we implement PWM as the primary method to regulate light output intensity for each color channel. PWM operates by rapidly switching the LED between on and off states, where the ratio of on-time to the total period (duty cycle) determines the perceived brightness. This provides linear brightness control and minimal color shifting across brightness levels. To address potential synchronization challenges between the three color channels, which could result in unwanted color artifacts, we utilize high-frequency PWM signals operating at 1 MHz. This PWM-based approach offers significant advantages over Digital-to-Analog Converter (DAC) implementation, including simpler circuit design, lower cost, and better reliability due to fewer components.

\begin{figure}[htp!]
    \centering
    \includegraphics[width=0.7\linewidth]{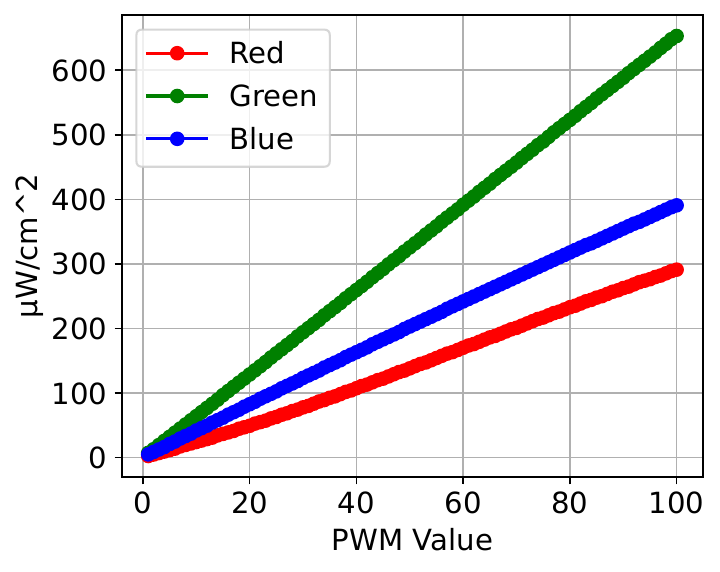}
    \caption{Optial power vs PWM values for different colors of RGB LED.}
    \label{fig:power-vs-pwm}
\end{figure}

The binary data is converted to symbols once the transmitter receives it. Given a symbol set $s_{1},...,s_{N}$ where $N$ is the number of total symbols, each symbol is determined as a constellation point in the CIE-1931 color space, as shown in Fig.~\ref{fig:constellation}. Each symbol is then converted to a PWM configuration that controls the current for the LED. To avoid optical flickering, the total optical power should remain constant, i.e., $P_{R}+P_{G}+P_{B}=1$, where $P$ is the light power. The specific PWM value for symbol $s_i$ can be calculated as:
\begin{equation}
\begin{bmatrix}
P_{R_i} \\ P_{G_i}
\\ P_{B_i}
\end{bmatrix} = \begin{bmatrix}
x_R & x_G & x_B \\
y_R & y_G & y_B \\
1 & 1 & 1 \\
\end{bmatrix}^{-1}\begin{bmatrix}
 x_i\\y_i
 \\1
\end{bmatrix},
\label{eq:csk}
\end{equation}
where $x_i$ and $y_i$ are the symbol's coordinates in the CIE color space, and ($x_R$, $y_R$), ($x_G$, $y_G$), ($x_B$, $y_B$) are the positions of red, green, and blue LEDs, respectively. In Eq.~(\ref{eq:csk}), it is assumed that the luminous intensity for different channels of the RGB LED is equivalent. In practical implementations, RGB LEDs may exhibit variations in color output due to inherent differences in the light efficiency of each LED channel. To compensate for these deviations, calibration coefficients are determined using a true color sensor. Figure~\ref{fig:power-vs-pwm} shows the optical power for different colors with varying PWM duty cycles. It is noted that the red LED has the lowest efficiency, while the green LED has the highest. To maintain the optical power constraint, the PWM values are multiplied by their respective efficiencies, which can be formulated as follows:
\begin{equation}
    \begin{bmatrix}
P_{R} \\ P_{G}
\\ P_{B}
\end{bmatrix}=\begin{bmatrix}
P_{R_i} \\ P_{G_i}
\\ P_{B_i}
\end{bmatrix} \cdot \begin{bmatrix}
e_{R} \\ e_{G}
\\ e_{B}
\end{bmatrix},
\end{equation}
where $e_{R}$ is 100\%, $e_{G}$ is 45\%, and $e_{B}$ is 75\%, which are acquired by using the red LED's optical power as the maximum power threshold.

\begin{figure}[htp!]
    \centering
    \includegraphics[width=0.8\linewidth]{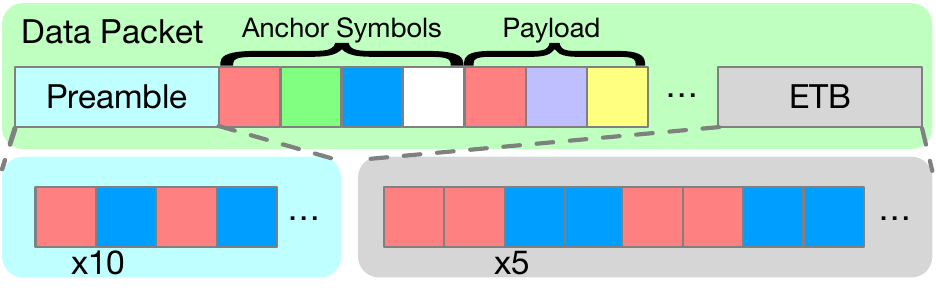}
    \caption{Format of the data packet.}
    \label{fig:packet-format}
\end{figure}

Aside from defining symbols, we also designed the packet format. As shown in Fig.~\ref{fig:packet-format}, data packets begin with preambles and conclude with an ``end of transmission block'' (ETB). Before transmitting the payload, anchor symbols are inserted into the signal. The preamble consists of ten consecutive transitions between two predefined symbols - in this case, two constellation points (red and blue) selected from 4-CSK. The ETB also employs color toggling, but with five transitions where each red and blue state persists for two time units.

\subsection{Receiver}
\label{sec:anchor}
Tandem solar cells are designed to enhance spectral coverage and response by incorporating multiple junctions with well-defined and complementary absorption characteristics~\cite{bett1999iii,yamaguchi2018review,snaith2013perovskites}. 
By leveraging the spectral selectivity of each layer, tandem cells not only optimize energy harvesting but also enable partial color discrimination, making them particularly well-suited for color-coded VLC applications.  
Since commercially available tandem solar cells are not yet widely accessible, we construct a prototype that replicates their spectral selectivity by integrating multiple solar cells made from different materials, each chosen for its distinct absorption properties, as illustrated in Fig.~\ref{fig:array} (b). By positioning these cells in close proximity on a wooden board, we minimize layout-induced variations while capturing a broader range of wavelengths. The cells are connected through resistors, and their voltage signals are collected as input for demodulation, leveraging the combined spectral responses to effectively differentiate color-encoded signals.

\begin{figure}[htp!]
    \centering
    \includegraphics[width=\linewidth]{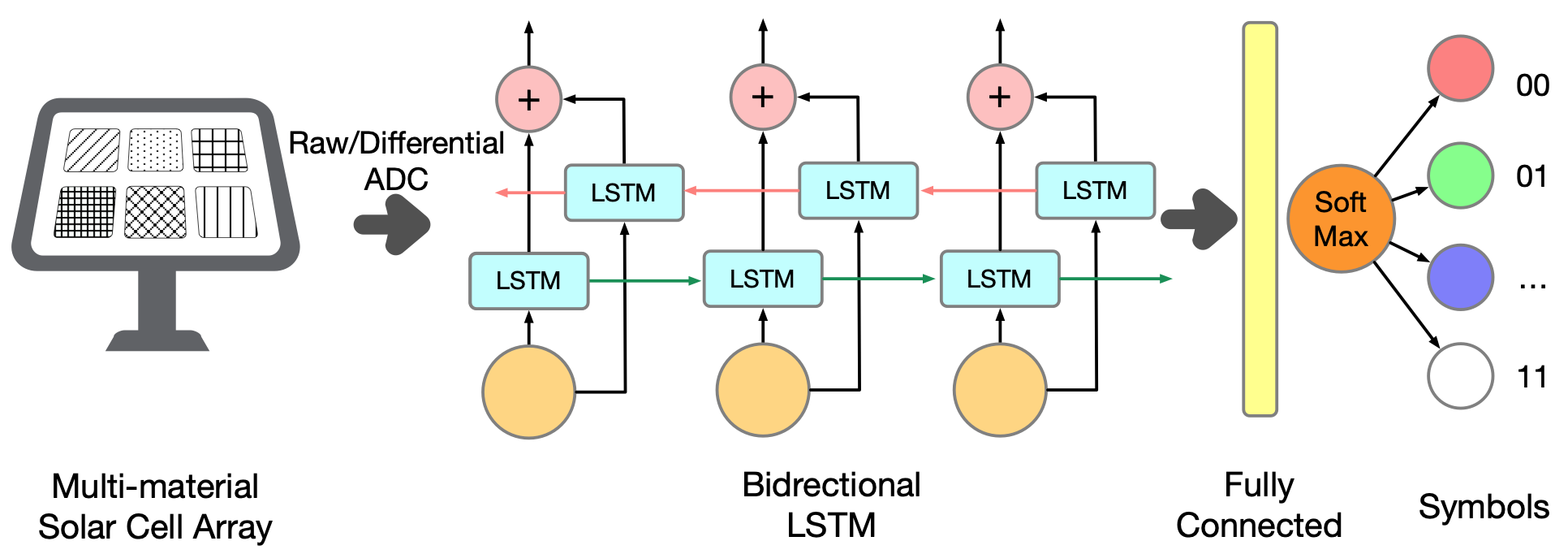}
    \caption{The demodulation pipeline for $N$-CSK employs a bidirectional LSTM followed by a fully connected layer to process photovoltaic signals. This architecture outputs probability distributions across the $N$ symbols defined in the $N$-CSK modulation scheme.}
    \label{fig:deocde}
\end{figure}
\subsection{Demodulation}

To demodulate CSK symbols from photovoltaic signals, we propose two methods: \textbf{Channel Estimation} and \textbf{Machine Learning (ML)}. For Channel Estimation, we adapt conventional signal processing techniques to the solar cell context. For ML, we design a bidirectional LSTM network and explore two training strategies: one using raw ADC samples from the solar cells (\textbf{ML without anchor}), and another using the difference between the raw samples and anchor measurements (\textbf{ML with anchor}).

\subsubsection{Channel Estimation:} This method follows the standard approach outlined in the IEEE CSK definitions, where the output of the solar cells can be modeled as  
    \begin{equation}
    y = Hs,
    \end{equation}
    where $y = [y_1, \ldots, y_N]^T \in \mathbb{R}^{N \times 1}$ represents the observed voltage signals from $N$ solar cells, $H \in \mathbb{R}^{N \times 3}$ is the VLC channel matrix that accounts for color crosstalk induced by imperfect filtering, and $s \in \mathbb{R}^{3 \times 1}$ denotes the power levels of the RGB channels.  

    In our case, with $N = 7$, the system is expressed as:  
    \begin{equation}
    \begin{bmatrix}
    h_{r1} & h_{g1} & h_{b1} \\
    h_{r2} & h_{g2} & h_{b2} \\
    \vdots & \vdots & \vdots \\
    h_{r7} & h_{g7} & h_{b7} \\
    \end{bmatrix} 
    \begin{bmatrix}
    s_{R} \\
    s_{G} \\
    s_{B} \\
    \end{bmatrix} = 
    \begin{bmatrix}
    y_{1} \\
    y_{2} \\
    \vdots \\
    y_{7} \\
    \end{bmatrix},
    \label{eq:channelEstimation}
    \end{equation}
    
    where $h_{ri}$, $h_{gi}$, and $h_{bi}$ represent the response of the $i$-th solar cell to the R, G, and B channels, respectively, while $s_R$, $s_G$, and $s_B$ are the power levels of the three channels. The voltage outputs from the solar cells are given by $y_1, y_2, \ldots, y_7$.  

    The decoding process consists of two phases: a) \textbf{Calibration Phase:} The elements of the channel response matrix $H$ are estimated.
        b) \textbf{Decoding Phase:} Given the known response matrix $H$ and measured signals $y$, we apply least squares estimation to recover the transmitted signal $s$. The decoded symbols are determined based on the shortest Euclidean distance.  

\subsubsection{Machine Learning:}
Recently, ML-based methods~\cite{li2021nelora,li2024chirptransformer,sankhe2019oracle} have gained significant traction for wireless signal decoding and demodulation tasks. Unlike conventional demodulation techniques that rely on explicit mathematical models, ML approaches can adaptively learn the nonlinear relationships between received signals and transmitted symbols without requiring perfect channel knowledge, resulting in improved performance.
These methods fall into two different types: (1) ML is used to estimate the channel and then the conventional `matrix inverse' methods are used to compensate the channel and obtain the symbol, (2) End-to-end decoding: the neural network is trained to directly output the decoded symbol from the input symbol without having to first estimate the channel, i.e.,  the channel estimation and decoding are combined in a single step. Compared to the first type, the end-to-end decoding approach offers better performance by eliminating error propagation between separate estimation and decoding steps. It also effectively handles non-linear channel distortions and features that are difficult to model using linear matrix operations.
Therefore, we propose an end-to-end demodulation pipeline, specifically employing a bidirectional Long short-term memory (LSTM)-based neural network for sequence classification. This architecture captures contextual dependencies from both past and future time steps, improving pattern recognition in the received signals. Bidirectional processing is particularly beneficial for solar cell signals, where both preceding and subsequent information contribute to accurate classification. Additionally, LSTM's ability to model long-range dependencies helps address the temporal complexities inherent in solar cell measurements.

The network consists of two primary components: a bidirectional LSTM encoder and a fully connected classifier. Fig.~\ref{fig:deocde} illustrates the overall architecture of our proposed model. The input sequence $X \in \mathbb{R}^{B \times T \times D}$ is processed through a deep bidirectional LSTM, where $B$ is the batch size, $T$ is the sequence length, and $D$ is the input feature dimension. The LSTM encoder is defined as:
\begin{equation}
h_{t} = \text{BiLSTM}(x_{t}, h_{t-1}),
\end{equation}
where $h_t$ represents the hidden state at time step $t$.

The bidirectional LSTM consists of two layers with 64 hidden units in each direction, yielding a combined hidden representation of size $2H = 128$ per time step due to the concatenation of forward and backward states. To mitigate overfitting, we apply dropout with a probability of 0.2 between LSTM layers.  

The final hidden states from both directions are concatenated to form the sequence representation:
\begin{equation}
h_{\text{final}} = [h_{\text{forward}}; h_{\text{backward}}] \in \mathbb{R}^{2H}.
\end{equation}
This representation is then passed through a classification head comprising two fully connected layers with ReLU activation and dropout:
\begin{align*}
h_{\text{hidden}} &= W_1 h_{\text{final}} + b_1, \\
h_{\text{drop}} &= \text{Dropout}(\text{ReLU}(h_{\text{hidden}}), p=0.2), \\
z &= W_2 h_{\text{drop}} + b_2,
\end{align*}
where $W_{1} \in \mathbb{R}^{H \times 2H}$, $W_{2} \in \mathbb{R}^{C \times H}$, and $C$ denotes the number of symbols. The final output $z \in \mathbb{R}^{C}$ represents the probability distribution over the possible symbols.  

The network is trained using the Adam optimizer with an initial learning rate of $10^{-3}$. We implement an adaptive learning rate schedule with a reduction factor of 0.5 and a patience of 5 epochs, monitoring the validation loss for adjustment. The objective function is the standard cross-entropy loss:
\begin{equation}
\mathcal{L} = -\sum_{i=1}^{C} y_i \log\left(\frac{e^{z_i}}{\sum_{j=1}^{C} e^{z_j}}\right),
\end{equation}
where $y_i$ represents the true class labels and $z_i$ the predicted logits. The model is implemented using PyTorch.

\begin{figure}[htp!]
    \centering
    \includegraphics[width=0.65\linewidth]{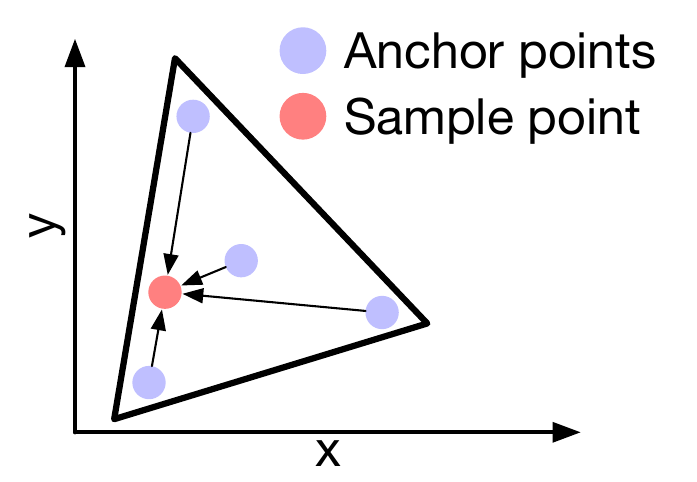}
    \caption{ Differential differences between the sample point and anchor points.}
    \label{fig:c}
\end{figure}

To accurately decode symbols from photovoltaic responses in tandem solar cells, a key challenge lies in managing channel variations caused by changes in distance and ambient light. These variations alter signal characteristics under different operational conditions, potentially compromising decoding accuracy. To mitigate these effects, we propose Differential Shift from Anchors:

  As illustrated in Fig.~\ref{fig:c}, we propose using the differential shift from anchor symbols as features for classification. Specifically, given a set of observed data points $d_1, d_2, \ldots, d_n$, we compute their differences relative to a predefined set of anchor symbols $a_1, a_2, \ldots, a_m$. Here, $d_n$ represents the $n$-th sample to be demodulated, while $a_m$ corresponds to the signal received when transmitting the $m$-th anchor symbol.  

    The differential shift $\Delta_{i,j}$ between sample $d_i$ and anchor $a_j$ is defined as:
    \begin{equation}
    \Delta_{i,j} = f(d_i, a_j),
    \end{equation}
    where $f(\cdot)$ is a difference function that can take various forms, such as:
    \begin{equation}
    f_{\text{abs}}(d_i, a_j) = \frac{|d_i - a_j|}{a_j}.
    \end{equation}

    This computation results in a feature vector for each sample:
    \begin{equation}
    \boldsymbol{\Delta}_i = [\Delta_{i,1}, \Delta_{i,2}, \ldots, \Delta_{i,m}].
    \end{equation}

    Compared to \textbf{Channel Estimation} method and the ML \textbf{using the raw output from solar cells}, the \textbf{ML with anchor} approach offers superior resilience to channel variability by leveraging a key advantage: anchor symbols and data symbols experience identical channel distortions. This preservation of differential relationships eliminates the need for explicit channel estimation and computationally intensive retraining procedures such as few shot learning. Consequently, this method provides a mathematically elegant and operationally robust solution, enabling immediate and reliable functioning in new environments, such as different distances or ambient lighting,  without the delays typically associated with parameter estimation or model fine-tuning.

\section{Implementation}
\label{s:impl}
\begin{figure}[htp!]
    \centering
    \includegraphics[width=0.6\columnwidth]{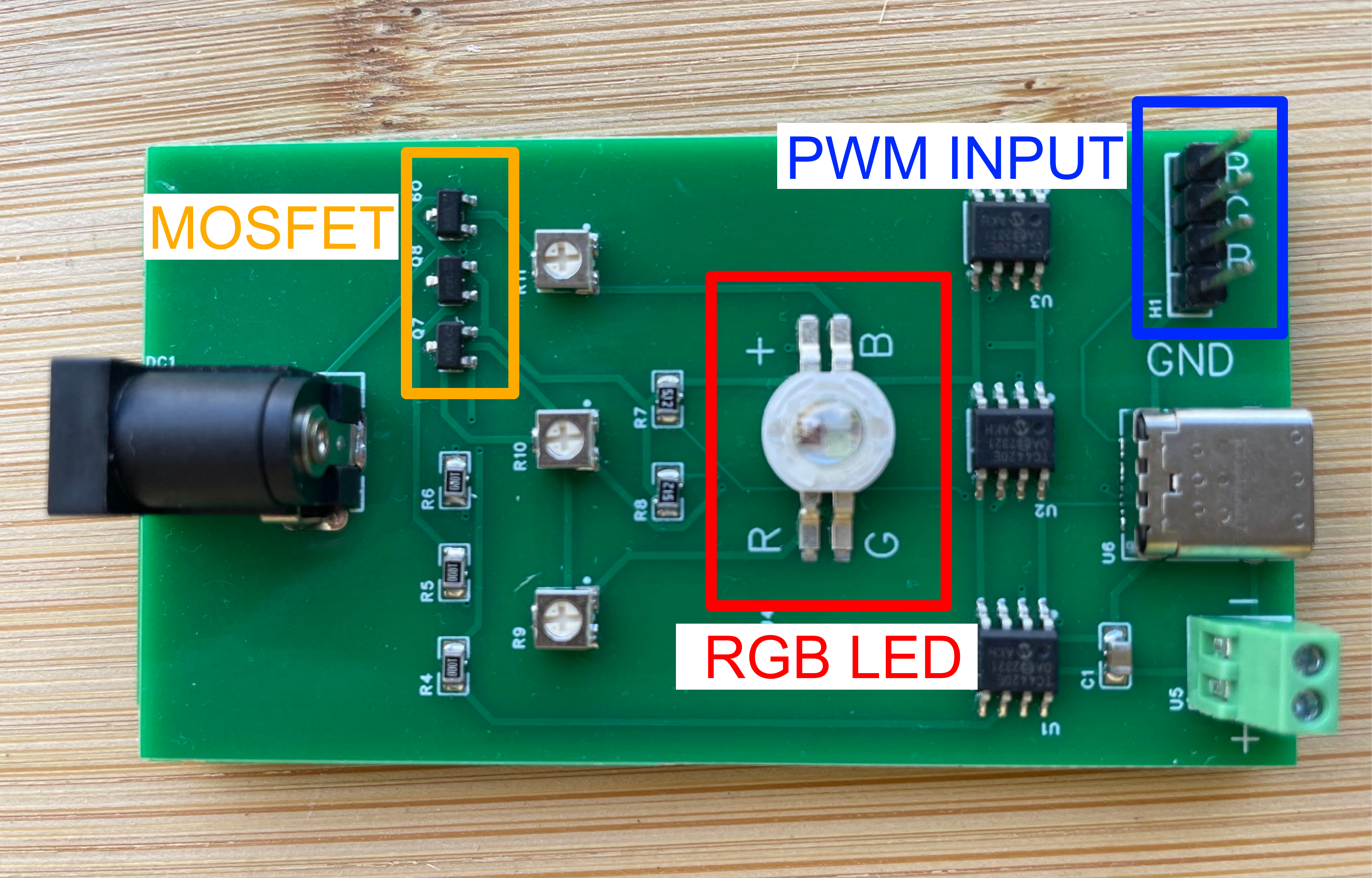}
    \caption{\cname transmitter.}
    \label{fig:transmitter}
\end{figure}

\textbf{Transmitter.} As illustrated in Figure~\ref{fig:transmitter}, our system employs a commercial RGB LED (model FD-3RGB-Y2)~\cite{led} as the primary illumination source. This LED emits three distinct color bands centered at wavelengths of 625 nm (red), 525 nm (green), and 465 nm (blue), enabling precise spectral control across the visible spectrum.  

To facilitate rapid and reliable switching between the on and off states, the LED is driven by an AO3400A MOSFET~\cite{mosfet}, which functions as a high-speed electronic switch with minimal switching losses and low on-state resistance. For precise temporal control of illumination, we utilize an Arduino DUE microcontroller, which features an ARM Cortex-M3 processor operating at 84 MHz.  

The Arduino DUE generates three independent PWM signals, each dedicated to modulating the intensity of a specific color band with high temporal resolution. These PWM signals are directly connected to the MOSFET gate terminals ($V_{in}$), allowing independent intensity modulation of each color channel through duty cycle adjustment. This configuration ensures precise control over the spectral composition of the output light while maintaining excellent temporal stability and reproducibility.

\textbf{Receiver.} The receiver consists of a solar cell array designed to mimic the behavior of seven tandem solar cells, each fabricated from different materials while maintaining consistent dimensional specifications across the assembly. Each solar cell is connected to a resistor, and the voltage signal across the resistor is amplified using an external circuit. To ensure uniform sensitivity and response characteristics, the gain values are individually calibrated and adjusted for each solar cell.
\begin{figure}[htp!]
    \begin{subfigure}[b]{0.48\columnwidth}
    \centering
    \includegraphics[width=\linewidth]{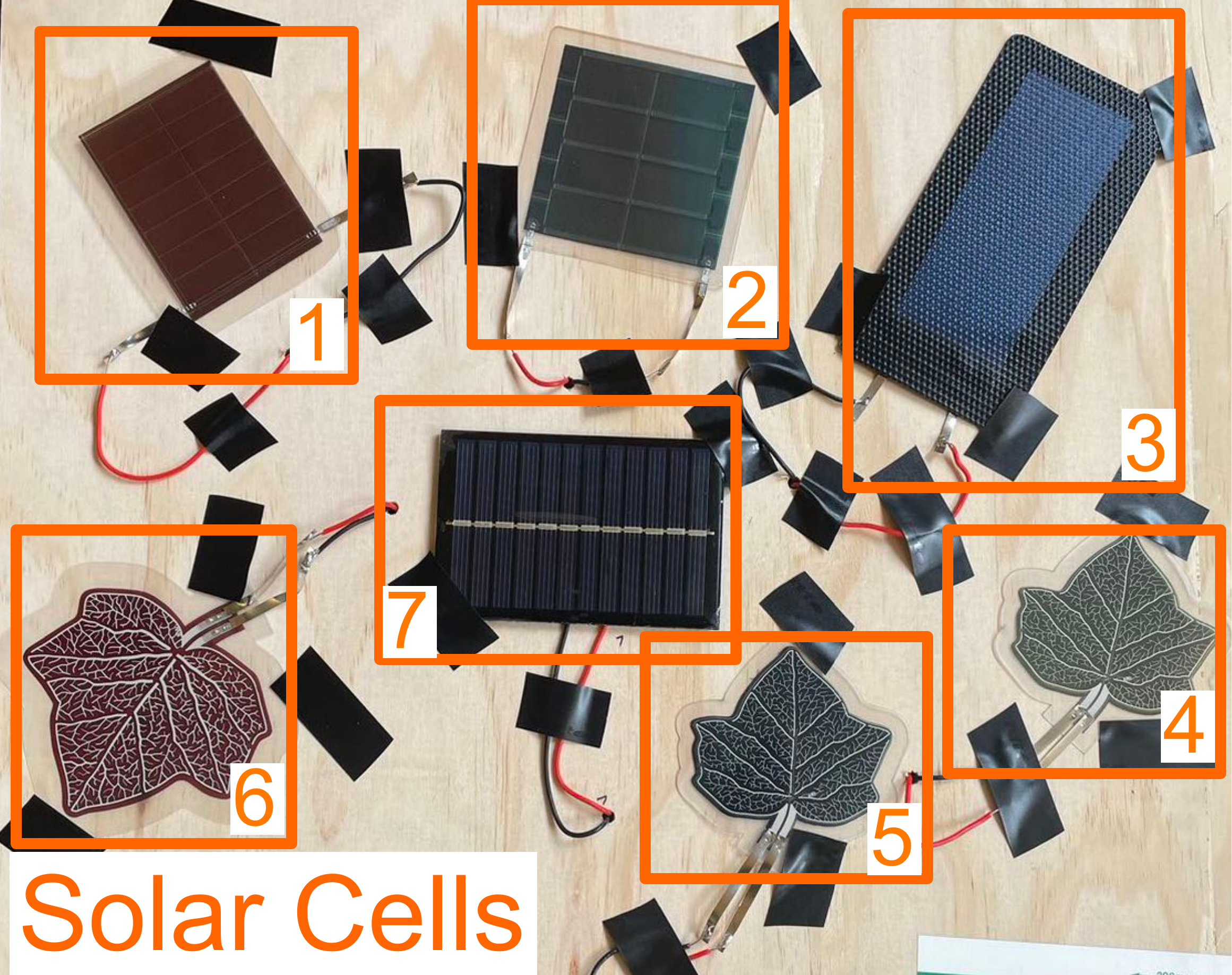}
    \caption{Front}
    \label{fig:receiver}
    \end{subfigure}
    \hfill
    \begin{subfigure}[b]{0.48\columnwidth}
    \centering
    \includegraphics[width=\textwidth]{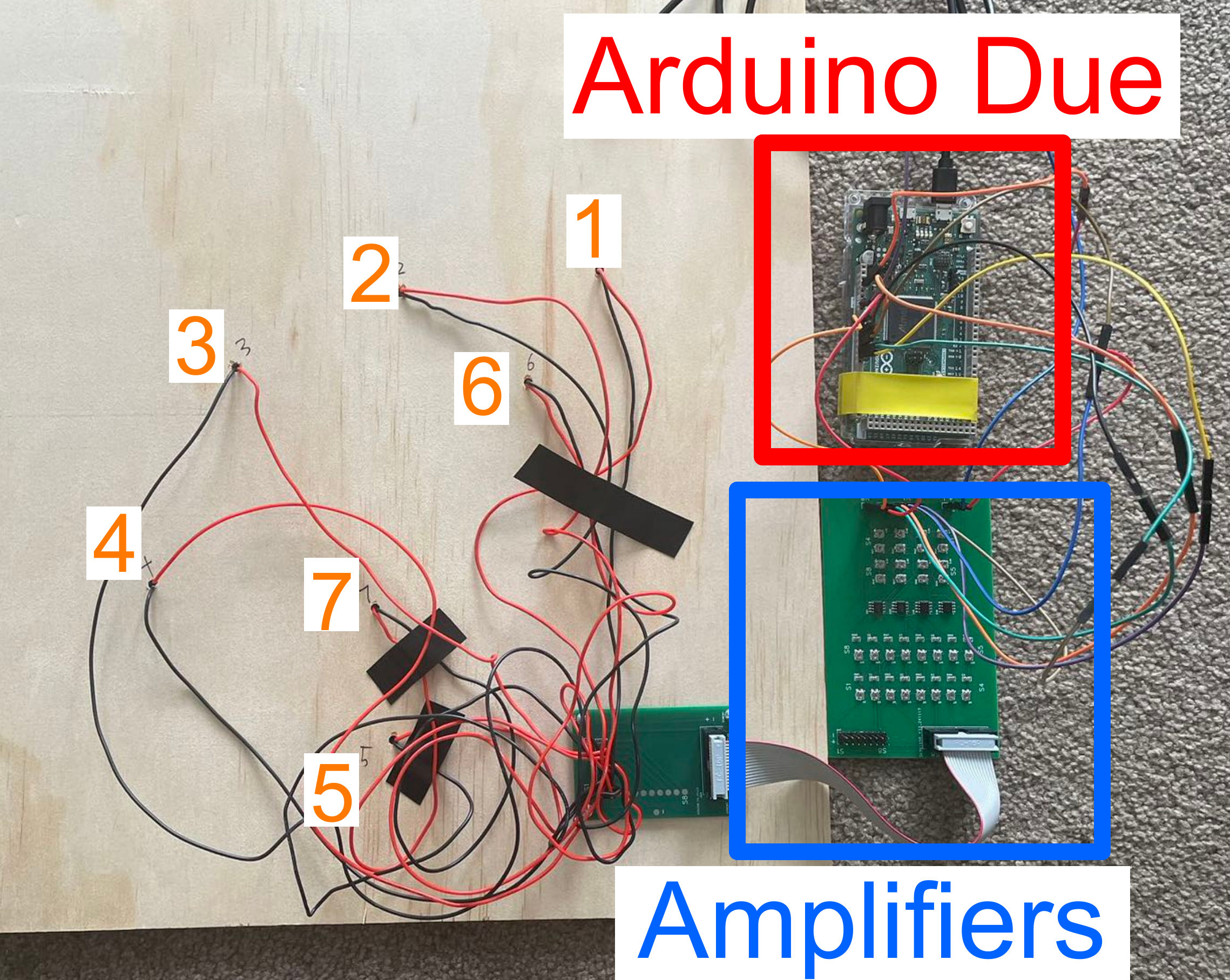}
    \caption{Back}
    \end{subfigure}
    \caption{\cname receiver.}
\end{figure}
The voltage signals are digitized using the integrated analog-to-digital converter (ADC) within the Arduino Due micro controller platform, which offers a high-precision 12-bit resolution for accurate signal quantization. To ensure synchronized data acquisition, the sampling rate is specifically configured to be integer multiples of both the transmitter's baud rate and refresh rate, thereby minimizing timing-related artifacts and maintaining signal integrity throughout the acquisition process. The acquired sensor data is efficiently transmitted via the serial port to a laptop. This computer system is equipped with an Intel i5 processor and 32 GB of RAM.

\section{Evaluation}
\label{s:eval}
\begin{figure}[htp!]
    \centering
    \includegraphics[width=0.7\linewidth]{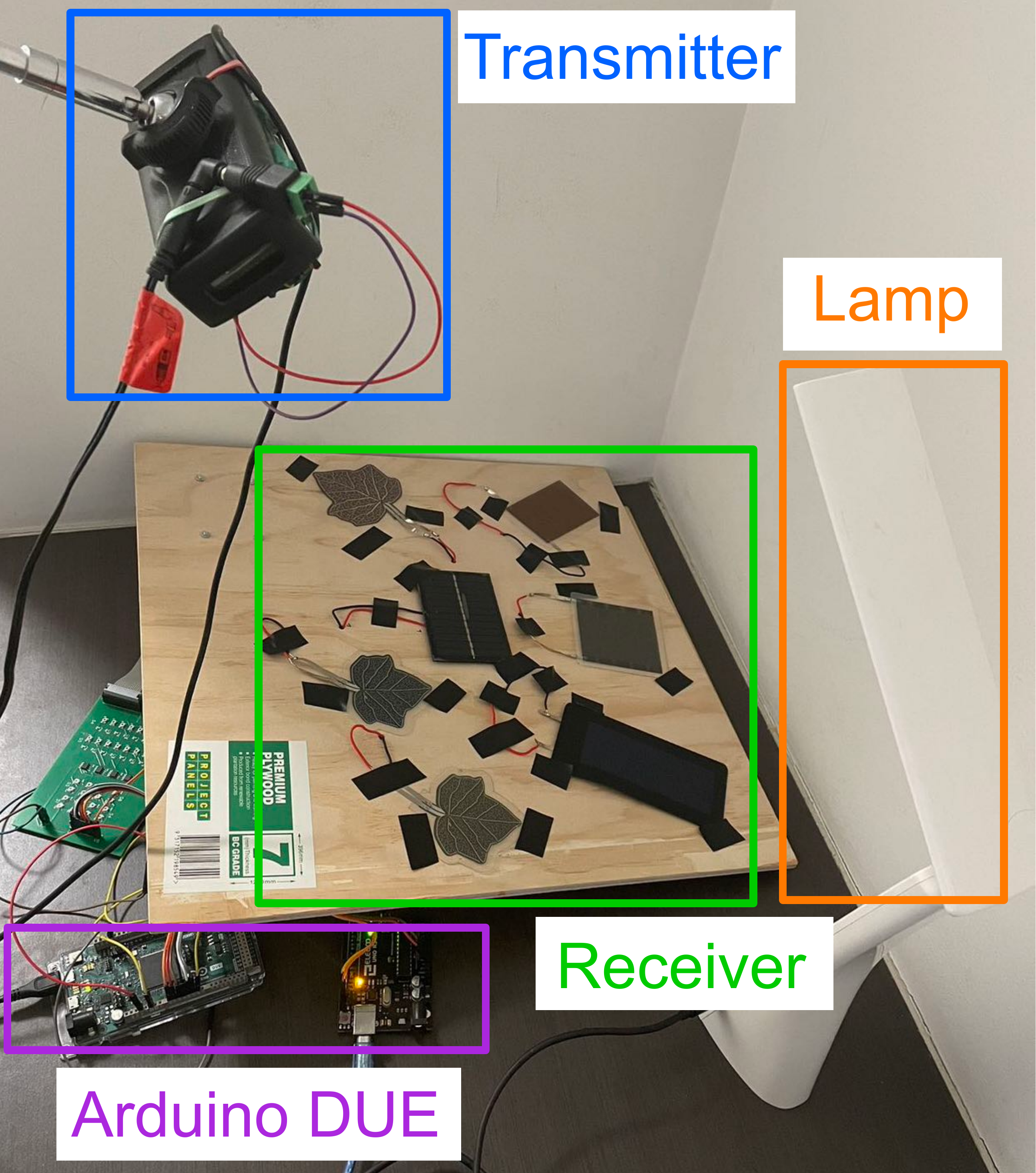}
    \caption{Experimental setup of \cname.}
    \label{fig:setup}
\end{figure}

To evaluate the performance of \cname, we collected data across multiple scenarios.  
Specifically, we utilized a tablet floor stand as the mounting apparatus, a RGB LED (model FD-3RGB-Y2)~\cite{led} as the controlled light transmitting source, and a white table lamp (model MTD4.5-M/K-19, 4.5W) as controllable ambient illumination. This setup, as shown in Fig.\ref{fig:setup}, allowed us to systematically regulate the distance between the transmitter and receiver, transmitter color, as well as the ambient light conditions, ensuring consistent experimental conditions. Details of the dataset are summarized in Table~\ref{tab:controlled}.  

\begin{table}[htp!]
\centering
\caption{Experimental dataset parameters.}
\label{tab:controlled}
\begin{tabular}{cc} 
\toprule
\multicolumn{1}{c}{Variable} & Values (Hz/Lux/cm) \\ 
\midrule
Baud Rate (25 cm, 0 Lux) & \begin{tabular}[c]{@{}c@{}}50, 100, 200, 300, 400, 500, \\ 600, 700, 800, 900, 1000\end{tabular} \\ 
\midrule
Ambient Light (25 cm, 500 Hz) & 0, 147, 618, 1154 \\ 
\midrule
Distance (0 Lux, 500 Hz) & 25, 30, 35, 40, 45, 50 \\ 
\bottomrule
\end{tabular}
\end{table}

\subsection{Sustainable Baud Rate Determination}

\begin{figure}[htp!]
    \centering
    \includegraphics[width=0.75\linewidth]{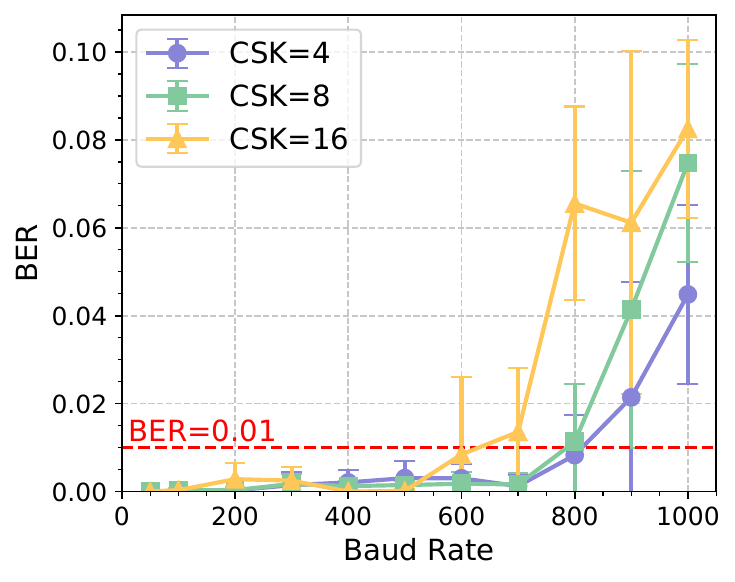}
    \caption{Baud rate vs. BER.}
    \label{fig:overall}
\end{figure}
This section details the experimental determination of the maximum reliable baud rate for the \cname prototype. The transmitter–receiver distance was fixed at 25 cm, and experiments were performed under dark conditions (0 Lux). The transmitter's baud rate was varied from 50 Hz to 1000 Hz, while all solar cells were sampled at 2 kHz using the Arduino ADC.

For signal transmission, the ASCII-encoded word `hello' was used, represented in binary as follows: \begin{center} 01101000 01100101 01101100 01101100 01101111 \end{center}

The `hello' bit sequence was transmitted 100 times. To ensure statistical robustness, a five-fold cross-validation was employed, with the average bit error rate (BER) and its standard deviation reported in Fig.~\ref{fig:overall}. The results show that \cname maintains a low BER (below 1\%) for baud rates up to 500 Hz. However, beyond this threshold, the BER increases significantly, particularly for CSK-16, as shorter symbol intervals make it increasingly difficult to distinguish among multiple symbols. We, therefore, use 500 Hz as the default baud rate for subsequent evaluations.

\begin{figure}[htp!]
    \centering
    \includegraphics[width=0.75\linewidth]{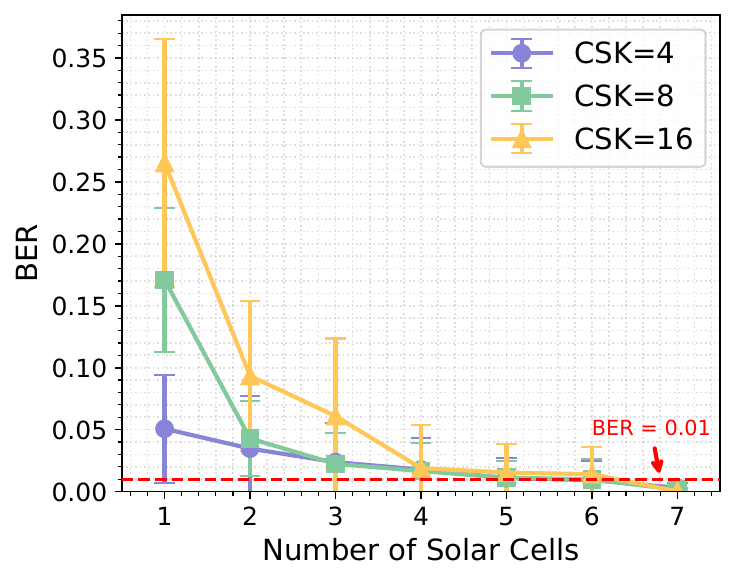}
    \caption{Number of cells vs. BER.}
    \label{fig:nc}
\end{figure}

\subsection{Impact of the Number of Solar Cells}

Following the determination of the optimal baud rate at 500 Hz, we conducted a detailed investigation into the relationship between the number of solar cells and system performance. To systematically evaluate this effect, we employed a random sampling approach, selecting varying numbers of solar cells from our array. For each configuration, we measured and computed the average BER along with its standard deviation, as illustrated in Fig.~\ref{fig:nc}.  

Our experimental results indicate a strong positive correlation between the number of solar cells and decoding accuracy. Notably, when utilizing a configuration of 7 solar cells, the system achieved a significantly improved BER below $10^{-2}$ even with 16-CSK, demonstrating that increasing the number of independent solar cell inputs enhances the reliability of the optical communication system. This performance improvement can be attributed to the increased spectral diversity and feature redundancy provided by multiple solar cells of varying absorption characteristics, which help mitigate the effects of local noise and interference. Subsequent evaluations are conducted with 7 solar cells.

\subsection{Impact of the Number of Anchors}

For 4-CSK at a distance of 45 cm, we systematically investigated the effect of varying the number of anchors on the demodulation performance, with results presented in Fig.~\ref{fig:45cm}.  We observed that beyond four anchors, the BER did not improve despite adding more anchors. This finding underscores the fundamental constraints of the anchor-based approach in scenarios where signal strength degradation becomes the dominant factor. 

Surprisingly, Fig.~\ref{fig:45cm} also reveals that the Channel Estimation method introduced in Section~\ref{sec:anchor} fails to improve the BER, even with additional information from the anchor points. We hypothesize that this is due to the shortest Euclidean distance criterion used in the decoding process, which relies on static feature comparisons and does not effectively capture the temporal dependencies inherent in solar cell measurements, e.g., due to response time, flickering, or gradual transitions between symbol states. This observation further justifies our design choice of a bidirectional LSTM-based decoder.  Additionally, we note that using fewer than three anchor points renders Eq.~(\ref{eq:channelEstimation}) underdetermined during the calibration phase. Consequently, no results are reported in Fig.~\ref{fig:45cm} for cases where the number of anchors is one or two.

Based on our comprehensive evaluation of the trade-off between computational complexity and performance improvement, we determined that using four anchors with Differential Shift from Anchors approach provides an optimal balance for our application. This configuration was subsequently adopted as the standard configuration for all further experiments in this study.

\subsection{Impact of Anchor Selection}

Following our previous experiments, we further investigated the effect of anchor selection on system performance. For this study, we used the model trained with data collected at 25 cm and tested it with data from 45 cm.  

\begin{figure*}[t]
    \centering
    \begin{minipage}[t]{0.32\linewidth}
        \centering
        \includegraphics[width=\textwidth]{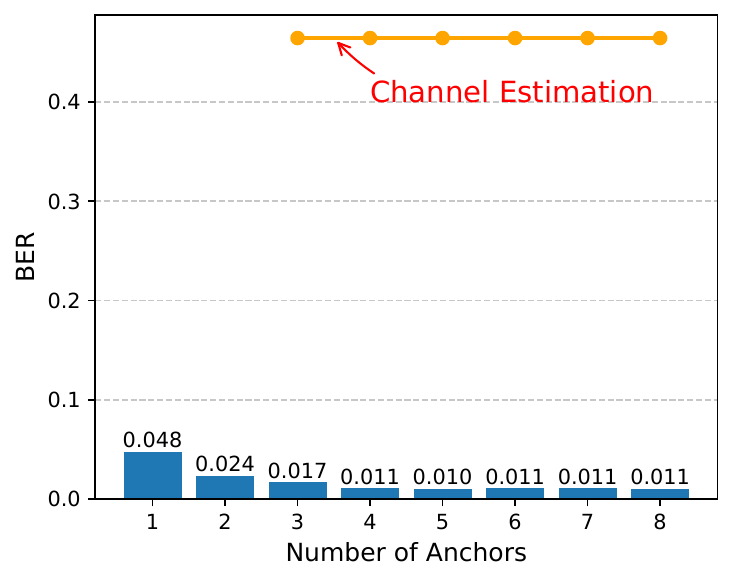}
        \caption{BER vs. \# of anchors.}
    \label{fig:45cm}
    \end{minipage}%
    \hfill%
    \begin{minipage}[t]{0.32\linewidth}
        \centering
        \includegraphics[width=0.8\textwidth]{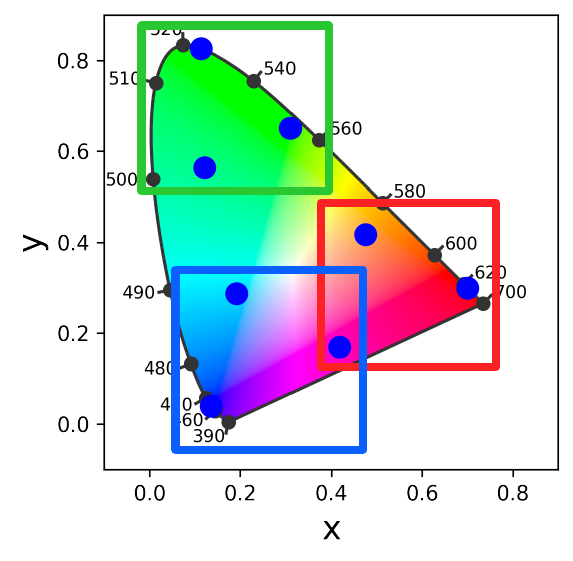}
        \caption{Three groups of anchors.}
        \label{fig:group}
    \end{minipage}%
    \hfill%
    \begin{minipage}[t]{0.32\linewidth}
        \centering
        \includegraphics[width=\textwidth]{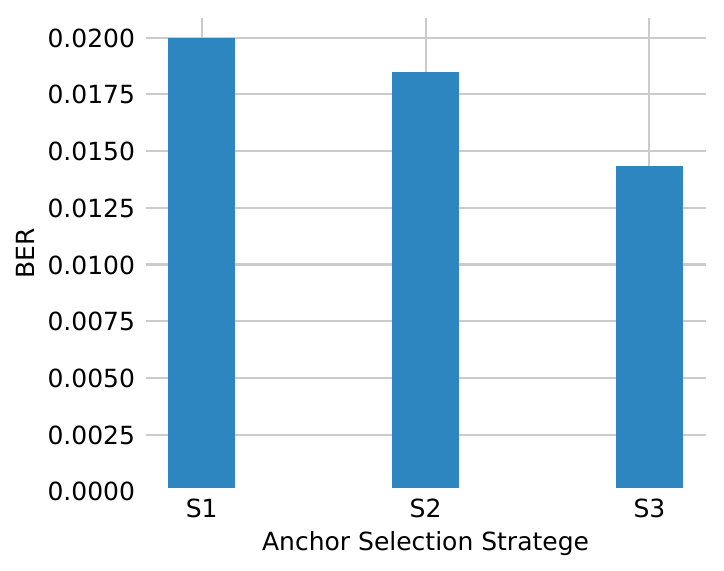}
        \caption{BER vs. anchor selection.}
        \label{fig:ber_group}
    \end{minipage}
\end{figure*}
As illustrated in Fig.~\ref{fig:group}, we selected eight anchor candidates (depicted as blue points), corresponding to the constellation points of 8-CSK. These candidates were grouped into three distinct sets, indicated by different colored rectangles (red, blue, and green). We then evaluated three anchor selection strategies:  
\begin{itemize}
    \item (S1) Selecting three anchors from a single group,  
    \item (S2) Selecting two anchors from one group and one from another,  
    \item (S3) Selecting one anchor from each group.  
\end{itemize}  
It is noted that selecting the same point multiple times will be excluded from the strategies. The average anchor distances for these strategies were 0.2662, 0.4295, and 0.4858, respectively. The corresponding BER values are presented in Fig.~\ref{fig:ber_group}, showing that S3 achieved the lowest BER, while S1 resulted in the highest.  
These results indicate that anchor selection should prioritize maximizing the distance between selected points, as greater separation enhances differentiation, similar to the principles used in constellation point selection.

\subsection{Evaluation on Unseen Channels}
In this section, we evaluate \cname with 4 anchors and 7 solar cells under \textit{unseen} channels, where channel variations result from changes in the Tx-Rx distance or ambient light intensity (lux). To achieve this, we apply leave-one-out training for the LSTM, training it on data from specific distances or lux levels while leaving one out for testing. 

\subsubsection{Channel Changes with Distance}

\begin{figure*}[t]
 \vspace{-3mm}
    \centering
    \includegraphics[width=\linewidth]{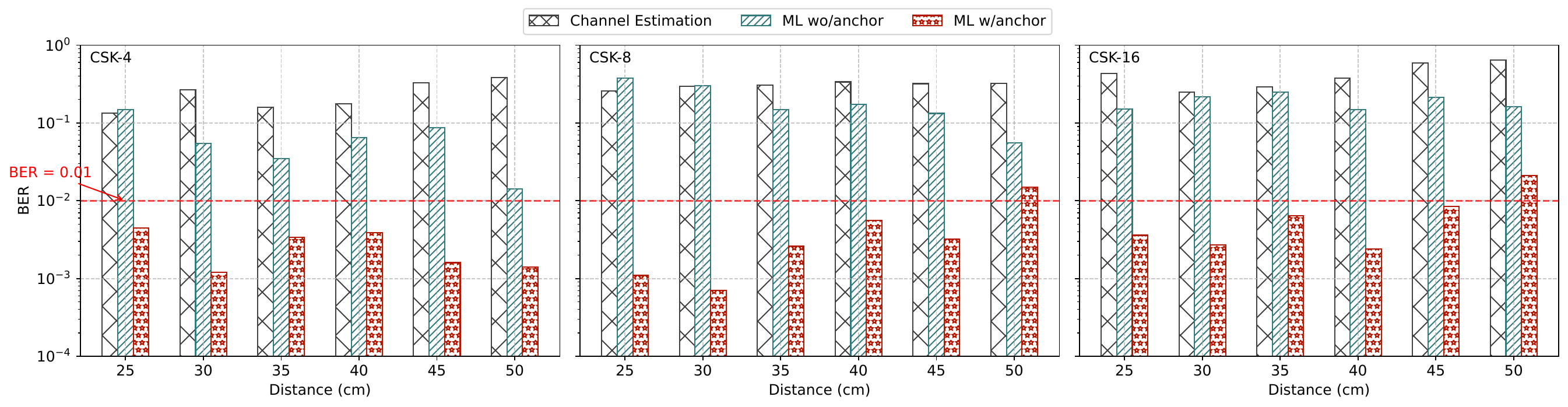}
    \vspace{-6mm}
    \caption{BER as a function of distance.}
    \label{fig:distance}
\end{figure*}

As distance increases, channel characteristics vary. In our system, light intensity diminishes with distance and the relative spacing between solar cells shifts, further affecting signal reception. To address these variations, we evaluated the three approaches discussed in Section~\ref{sec:anchor}: calibrating signals using a classical channel estimation technique (Channel Estimation), feeding raw solar ADC data into an LSTM (ML without anchors), and providing differential ADC input to an LSTM that leverages anchor information (ML with anchors). Data were collected from six different distances, and the leave-one-out evaluation results are shown in Fig. \ref{fig:distance}. 

We can see that, ML-based methods generally outperform Channel Estimation, with ML using anchors significantly surpassing Channel Estimation in all cases. These results demonstrate the effectiveness of our bidirectional LSTM network and the differential ADC input method with anchor information for generalizing \cname across unseen channels. Furthermore, the proposed ML method requires no retraining or fine-tuning, reducing computational overhead, enabling immediate deployment in diverse environments, minimizing storage requirements, and ensuring consistent performance without environment-specific calibration, a significant advantage for resource-constrained systems.

\subsubsection{Channel Changes with Ambient Light}
Since our method encodes information via color rather than amplitude, it is more susceptible to ambient light interference. In our setup, the transmitted light interacts with ambient illumination from a standard white LED table lamp. We observed that ambient light induces a "color pulling" effect, shifting perceived colors from their intended positions in the color space. To evaluate system performance under varying conditions, we conducted a leave-one-out experiment across four ambient lighting scenarios representative of typical industrial and commercial environments: dark room (0 Lux), warehouse lighting (147 Lux), standard office illumination (618 Lux), and mechanical workshop conditions (1,154 Lux), following standard illuminance guidelines~\cite{toolbox2004illuminance}.

\begin{figure*}[htb!]
    \centering
    \includegraphics[width=\linewidth]{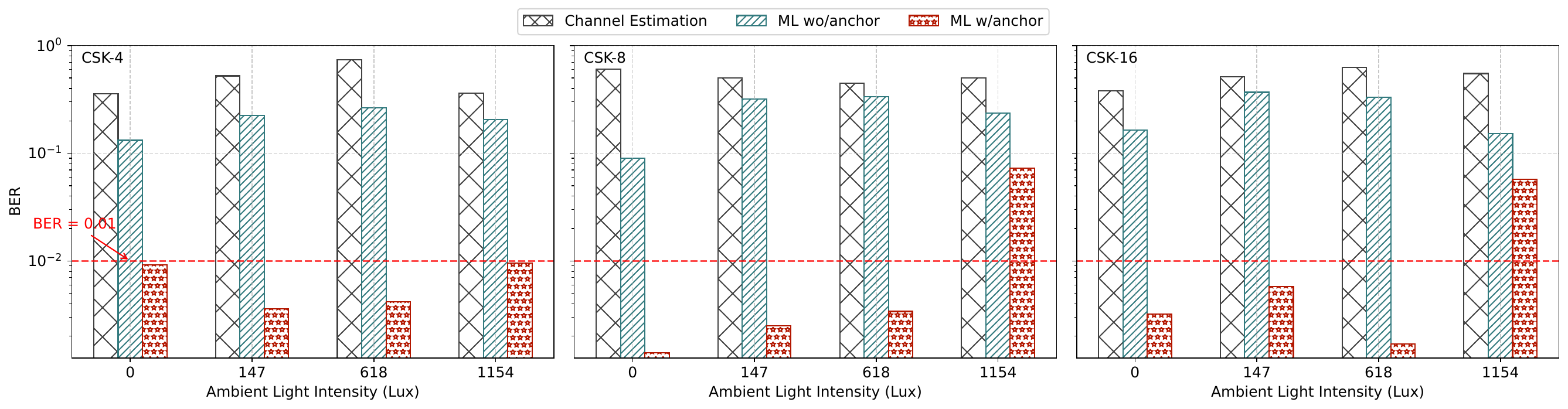}
     \vspace{-6mm}
    \caption{BER as a function of ambient light intensity.}
    \label{fig:diff-lux}
\end{figure*}

Similar to the distance experiments, three approaches are compared in Fig. \ref{fig:diff-lux}. The results show that ML using anchors significantly outperforms both Channel Estimation and ML without anchors in all cases, confirming its superior generalization across scenarios affected by distance or ambient light.

The most challenging scenario emerged under the intense lighting conditions of mechanical workshop illumination (1,154 Lux). At this high illuminance level, only the CSK-4 configuration maintained operational viability, primarily due to its more widely separated symbol constellation. Both CSK-8 and CSK-16 configurations exhibited significant performance degradation due to severe ambient light interference. However, even in this demanding environment, the ML with anchors approach demonstrated meaningful improvements, reducing the BER by up to 85\% compared to conventional methods.

\section{Discussion and Limitations}

\textbf{ADC Sampling Rate.} In the implementation of \cname, we adopted a default sampling rate of 2 kHz. However, this relatively modest frequency imposes constraints on the maximum baud rate the system can effectively support. According to the Nyquist-Shannon sampling theorem, accurate signal reconstruction requires a sampling rate at least twice the highest signal frequency component. Consequently, our current 2 kHz sampling rate theoretically limits the maximum baud rate to approximately 1 kbps, which may be insufficient for applications demanding higher data throughput. Furthermore, this limitation could adversely affect performance in noisy environments, where a higher sampling rate would enable more robust error correction and advanced signal processing techniques.

\textbf{Optimized Solar Cells.} Instead of relying on commercial solar cells, there is significant potential for developing specialized solar cells tailored to the dual functions of communication and energy harvesting. By strategically modifying semiconductor materials and photovoltaic compositions, we can achieve both more diverse spectral absorption characteristics for finer color differentiation and enhanced power conversion efficiency simultaneously. 
In the current implementation of \cname, the system operates with a maximum baud rate of 500 Hz, representing the upper frequency at which the solar cells can effectively detect and respond to color transitions. Optimizing these materials may significantly enhance both the response time and energy harvesting capabilities by enabling coverage of a wider spectral range, allowing simultaneous improvements in communication speed and energy conversion efficiency.

\textbf{Trade-off between Energy Harvesting and Communication.} When solar cells are used for both communication and energy harvesting, balancing these two functions becomes a critical design consideration. Increasing the duration allocated to energy harvesting enhances power sustainability but reduces communication throughput and reliability. Conversely, prioritizing communication improves data transfer rates at the expense of energy self-sufficiency. In this work, we emphasize the communication aspect of the system, but further exploration of this trade-off is warranted.  

Several strategies could address this challenge. One approach involves allocating distinct time intervals, dedicating specific periods exclusively to energy harvesting while reserving others for communication to optimize efficiency in both domains. Another approach entails dynamically adjusting the communication speed based on real-time task requirements while minimizing its impact on energy harvesting efficiency. Future work should explore these strategies to achieve an optimal balance between energy harvesting and communication performance in solar-assisted optical communication systems.

\section{Related Work}
\textit{Solar cells for VLC.}
While traditional RF signals have served as the backbone of mobile communications, VLC technology presents a compelling alternative with significant advantages. Most notably, VLC offers approximately 340 THz of license-free bandwidth and remains immune to electromagnetic interference, making it particularly valuable in environments where electromagnetic compatibility is crucial, such as hospitals and industrial facilities. Furthermore, VLC provides an inherent layer of security in its physical architecture, as visible light cannot penetrate solid walls. Additionally, VLC leverages existing lighting infrastructure that will be widely deployed in the future, resulting in reduced retrofitting costs compared to traditional communication systems. Driven by the growing adoption of sustainable Internet of Things (IoT) technologies\cite{nivzetic2020internet}, solar cells have recently garnered significant attention due to their dual functionality as both receivers and energy harvesters. This unique capability enables solar cells to simultaneously detect incoming VLC signals for data reception while harvesting light energy, making them an attractive component for energy-efficient VLC implementations. In the systems using solar cells as receivers, OOK modulation is widely adopted in the literature \cite{shin2016self} due to its simplicity and ease of implementation using low-cost off-the-shelf hardware components. However, OOK suffers from limited spectral efficiency as it can only transmit one bit per transition. To address this limitation, researchers have explored more advanced modulation schemes, such as Pulse-amplitude modulation (PAM) and Orthogonal frequency-division multiplexing (OFDM) \cite{wang2018using,kong2019toward}. While these methods have shown the improvement in the data rate, they primarily focus on controlling amplitude and flash frequency, resulting in limited spectral coverage due to their reliance on intensity modulation. Inspired by \cite{martinez2024evaluation, umetsu2019ehaas, wang2023spectral}, we observe that different solar cell materials exhibit unique spectral responses. These distinct characteristics enable the implementation of CSK in VLC systems using solar cell arrays as receivers—a novel approach that has not been previously demonstrated.

\textit{Color Shift Key Modulation.} The utilization of different colored lights to represent distinct symbols in the physical layer is comprehensively defined in the IEEE 802.15.7 standard~\cite{ieee2011ieee}. To optimize CSK modulation across various applications, researchers have made significant advances on both the transmitter and receiver sides of the communication system.
On the transmitter side, Martinez et al.~\cite{martinez2019design} introduced an innovative light-to-frequency converter (LTF) that eliminates the need for a DAC, thereby simplifying color control while maintaining system performance. This design requires only a single photodiode and ADC for signal decoding, significantly reducing system complexity. Furthermore, Monteiro et al.~\cite{monteiro2014design} enhanced the constellation point selection process by incorporating additional constraints, such as inter-channel crosstalk. Their approach can generate three-dimensional CSK constellations that can achieve specified luminaire output intensities while maintaining optimal communication performance.
On the receiver side, Hu et al.~\cite{hu2015colorbars} demonstrated the feasibility of using smartphone cameras as novel receivers for light signal detection, achieving higher data rates compared to traditional OOK methods. Additionally, Ye et al.~\cite{ye2021spiderweb} conducted an in-depth investigation of the color pulling effect, a phenomenon particularly relevant in scenarios where photodiodes are positioned beneath Organic Light-Emitting Diode (OLED) screens. To the best of our knowledge, \cname is the first attempt to use solar cells as receivers for decoding color-coded signals in VLC.

\section{Conclusion}
In this work, we explored the feasibility of using solar cells for color-based VLC. We developed \cname, a system that employs spectrally diverse solar cells and a machine learning-based decoder to receive CSK-modulated signals. Our results show that anchor-based differential measurements enhance robustness against channel variations, outperforming traditional channel estimation and calibration methods.  
Through extensive testing, we analyzed key factors such as baud rate, distance, anchor selection, and ambient light. \cname maintains reliable communication at up to 500 Hz and 45 cm, even with a modest ADC sampling rate of 2 kHz, while exhibiting strong resilience to ambient illumination changes. Our study positions \cname as a feasible alternative to photodiode-based CSK-VLC receivers, paving the way for sustainable and scalable VLC solution for the growing Internet of Things market. As solar cell technology advances, particularly with tandem solar cells, \cname can evolve to support higher data rates and broader applications.

\begin{acks}
This work was partly supported by the Australian Research Council Discovery Project DP210100904, UNSW Scientia PhD Scholarship Scheme and CSIRO Data61 PhD Scholarship Program.
\end{acks}

\bibliographystyle{ACM-Reference-Format}
\bibliography{reference}

\end{document}